\def\final{1}			

\ifdefined\preprint
  \documentclass[preprint,review,12pt]{elsarticle}
\fi
\ifdefined\final
  \documentclass[final,3p,times,twocolumn]{elsarticle}
\fi

\usepackage{graphicx,stfloats}
\usepackage{color}
\usepackage{pgf}
\usepackage{subcaption}
\usepackage[version=3]{mhchem}

\usepackage[nolist]{acronym}
\begin{acronym}
	\acro{DNS}{direct numerical simulation}
	\acroplural{DNS}[DNS]{direct numerical simulations}
	\acro{2D-DNS}{two-dimensional direct numerical simulation}
	\acroplural{2D-DNS}[2D-DNS]{two-dimensional direct numerical simulations}
	\acro{IFI}{intrinsic flame instability}
	\acroplural{IFI}[IFIs]{intrinsic flame instabilities}
	\acro{TDI}{thermodiffusive instability}
	\acroplural{TDI}[TDIs]{thermodiffusive instabilities}
	\acro{DLI}{Darrieus-Landau instability}
	\acroplural{DLI}[DLIs]{Darrieus-Landau instabilities}
	\acro{ITV}{Institute for Combustion Technology}
	\acro{H2}[\ce{H2}]{hydrogen}
	\acro{NH3}[\ce{NH3}]{ammonia}
	\acro{O2}[\ce{O2}]{oxygen}
	\acro{NOx}[NO\textsubscript{x}]{nitrogen oxide}
	\acroplural{NOx}[NO\textsubscript{x}]{nitrogen oxides}
\end{acronym}
\usepackage[capitalise]{cleveref}
\usepackage{longtable}
\usepackage{siunitx}
\newcommand{\ignore}[1]{}
\usepackage{xr}
\newcommand{\XHF}{X_{\rm H_2,F}}
\newcommand{\lF}{l_{\rm F}}

\newcommand{\rmd}{\mathrm{d}}
\newcommand{\Tu}{T_{\rm u}}

\graphicspath{{..}{Figures/Article}}

\biboptions{sort&compress}

\journal{XXXXXX}

\begin{document}

\begin{frontmatter}

\title{Comprehensive linear stability analysis for intrinsic instabilities in premixed ammonia/hydrogen/air flames}

\author[fir]{Terence Lehmann\corref{cor1}}

\author[fir]{Lukas Berger}
\author[fir]{Thomas L. Howarth}
\author[fir]{Michael Gauding}
\author[fir]{Sanket Girhe}
\author[sec]{Bassam B. Dally}
\author[fir]{Heinz Pitsch}

\address[fir]{Institute for Combustion Technology, RWTH Aachen University, Templergraben 64, 52056 Aachen, Germany}
\address[sec]{Clean Energy Research Platform, King Abdullah University of Science and Technology, Thuwal, 23955-6900, Saudi-Arabia}
\cortext[cor1]{Corresponding author: t.lehmann@itv.rwth-aachen.de}

\begin{abstract}
	
Two-dimensional direct numerical simulations of planar laminar premixed \acl{NH3}/\acl{H2}/air flames are conducted for a wide range of equivalence ratios, \ac{H2} fractions in the fuel blend, pressures, and unburned temperatures to study \acp{IFI} in the linear regime. For stoichiometric and lean mixtures at ambient conditions, a non-monotonic behavior of thermo-diffusive instabilities with increasing \ac{H2} fraction is observed. Strongest instabilities occur for molar \ac{H2} fractions of $40\%$. The analysis shows that this behavior is linked to the joint effect of variations of the effective Lewis number and Zeldovich number. \Acp{IFI} in \acl{NH3}/\acl{H2} blends further show a non-monotonic trend with respect to pressure, which is found to be linked to the chemistry of the hydroperoxyl radical \ce{HO2}. The addition of \ce{NH3} opens new reaction pathways for the consumption of \ce{HO2} resulting in a chain carrying behavior in contrast to its chain terminating nature in pure \ac{H2}/air flames. Theoretically derived dispersion relations can predict the non-monotonic behavior for lean conditions. However, these are found to be sensitive to the different methods for evaluating the Zeldovich number available in the literature.
	
\end{abstract}

\begin{keyword}
	
	ammonia \sep hydrogen \sep thermo-diffusive instability \sep linear stability analysis
	
\end{keyword}
\end{frontmatter}

\section*{Novelty and Significance Statement}
	
The novelty of this research is the systematic identification and explanation of a non-monotonic behavior of intrinsic flame instabilities \acp{IFI} in ammonia/hydrogen/air (\ce{NH3}/\ce{H2}/air) flames concerning the hydrogen content in the fuel and the pressure. To the author's knowledge, this study presents the largest parametric study for linear stability analyses of \ce{NH3}/\ce{H2}/air flames. Furthermore, a sensitivity analysis to the Zeldovich number is proposed to link the macroscopic effect of \acp{IFI} to the microscopic effects of chemistry. In light of possible applications of \ce{NH3} as zero-carbon fuel, this study is significant because the fundamental understanding of \acp{IFI} in \ce{NH3}/\ce{H2}/air flames is key for the analysis and modeling of such flames.

\acresetall
	
\section*{Author Contributions}

\textbf{TL}: Conceptualization, methodology, investigation, data curation, writing - original draft, writing - review \& editing
\textbf{LB}: Conceptualization, methodology, writing - review \& editing
\textbf{TLH}: Conceptualization, software, writing - review \& editing
\textbf{MG}: Conceptualization, supervision, writing - review \& editing
\textbf{SG}: Conceptualization, resources, writing - review \& editing
\textbf{BBD}: Conceptualization, supervision, funding acquisition, writing - review \& editing
\textbf{HP}: Conceptualization, supervision, funding acquisition, project administration, writing - review \& editing

\section{Introduction}
\label{sec:Introduction}

\Ac{H2} is a well-known carbon-free fuel candidate. The direct synthesis through electrolysis leads to high production efficiencies. However, its low volumetric energy density and associated storage conditions (i.e., pressures up to \SI{700}{\bar} or temperatures below \SI{20}{\kelvin}) make its transport difficult. One method to overcome these difficulties is by converting \ac{H2} to \ac{NH3}~\cite{ValeraMedina2018, Kobayashi2019, Elbaz2022}. Since \ac{NH3} transitions to its liquid state at around \SI{-33}{\degreeCelsius} and \SI{1}{\bar}, or \SI{20}{\degreeCelsius} and \SI{9}{\bar}, long distance transport and long duration storage are more viable when compared with \ac{H2}~\cite{Mueller2024}. \citet{Wijayanta2019} point out that the utilization of \ac{NH3} as \ac{H2} carrier instead of direct transportation of \ac{H2} even increases the overall efficiency and reduces costs. 

While premixed \ac{NH3}/air flames are characterized by low burning velocities and high ignition energies, they significantly improve when mixed with \ac{H2}~\cite{Kobayashi2019}. However, the introduction of \ac{H2} triggers \acp{IFI} driven by thermo-diffusive processes~\cite{Lee2010, Lee2010b, Ichikawa2015, Zitouni2023}. \Acp{IFI} can be separated into hydrodynamic, or Darrieus-Landau, instabilities (\acsu{DLI}), and \acp{TDI}~\cite{Law2000, Matalon2007}. While \ac{DLI} are caused by the density change across the flame, \ac{TDI} result from the disparity between thermal diffusivity and species diffusivities. This effect is characterized by non-unity Lewis numbers $\mathit{Le}_i$. In \ac{H2}/air flames, \acp{TDI} are known to cause a significant increase in total flame surface area and local burning rates~\cite{Berger2019, Berger2022b, Howarth2022}, and can impact the formation of \acp{NOx}~\cite{Wen2023}.

\acp{IFI} also occur for fuel blends of \ac{H2} and \ac{NH3} and have been experimentally observed in spherical expanding flames~\cite{Lee2010, Lee2010b, Ichikawa2015, Zitouni2023} and analyzed with respect to the mole fraction of \ac{H2}, $X_{\rm H_2}$, in the fuel mixture, defined as
\begin{equation}
	\XHF = \frac{X_{\rm H_2}}{X_{\rm H_2} + X_{\rm NH_3}}\,.
\end{equation}
\citet{Lee2010} investigated blends with $\XHF\leq 0.5$, showing a monotonic decrease of the Markstein number $\cal{M}$ with increasing $\XHF$, which is an indication for increased importance of \acp{TDI}. From this they concluded a generally dampening effect of \ac{NH3} on \ac{IFI}. \citet{Ichikawa2015} found a similar behavior of $\cal{M}$ for stoichiometric \ac{NH3}/\ac{H2}/air flames at ambient pressure. However, when further increasing the \ac{H2} mole fraction in the fuel, they observed $\cal{M}$ increasing again, indicating decreasing \acp{IFI}. For increasing pressure $p$ from \SI{1}{\bar} to \SI{5}{\bar}, $\cal{M}$ globally decreases, while the non-monotonic behavior becomes less pronounced. \citet{Zitouni2023} performed similar experiments for a wide range of equivalence ratios $\phi$ at ambient conditions, finding that the non-monotonic behavior of $\cal{M}$ is amplified with decreasing $\phi$.

A method to analyze the time and length scales of \acp{IFI} numerically is provided by the linear stability analysis of planar flames. This method focuses on the early onset of an instability, referred to as the linear regime, by applying a weak perturbation to the flame front. The perturbation will decay in a stable case, whereas a perturbation growth is observed in an unstable case. For sufficiently weak harmonic perturbations, the growth or decay will follow an exponential law with 
\begin{equation}
A(t)\propto\exp(\omega(k) t)\,, 
\end{equation}
\noindent
where $A(t)$ is the instantaneous perturbation amplitude at time $t$ and $\omega(k)$ is the growth rate. The particular dependence of $\omega$ on $k$, referred to as a dispersion relation, then depends on the conditions, such as fuel (blend), $\phi$, $T$, and $p$. With progressing development of the instability, the flame transitions to a non-linear and later on chaotic behavior, referred to as non-linear regime, where the flame cannot be assumed to be weakly perturbed. Note that the present study provides a linear stability analysis and, hence, only considers the linear regime.

For lean \ac{H2}/air flames, numerous studies on dispersion relations exist~\cite{Matalon2003, Altantzis2011, Altantzis2012, Berger2022}. \citet{Berger2022} showed that \acp{TDI} increase with increasing pressure and decreasing unburned temperature or $\phi$. In addition, they show that instabilities in \ac{H2}/air flames can be parameterized by a set of non-dimensional flame parameters~\cite{Berger2022}. Furthermore it has been shown that a close linking between the linear regime, described by dispersion relations, and the non-linear regime of fully developed instabilities exists. More specifically, the most prominent length scales in the non-linear regime coincide with the peak growth rates in the linear regime~\cite{Berger2019}. Also, the global consumption speed enhancement in developed flames correlates with the peak growth rate in dispersion relations~\cite{Berger2022b}.

Recently, \citet{Gaucherand2023} computed dispersion relations for \ac{NH3}/\ac{H2}/ air flames. They investigated fuel blends with $0.4 \leq \XHF \leq 1.0$ at atmospheric temperature and pressure for equivalence ratios of $\phi = 0.4, 0.5, \mathrm{~and~} 1.0$, as well as at elevated pressure of $p=10~\mathrm{bar}$ for $\phi = 0.5$. The presented results suggest a monotonic decrease of instabilities with decreasing $\XHF$, which is not in agreement with the findings of \citet{Ichikawa2015} and \citet{Zitouni2023}. However, the study applies a reduced chemical kinetic model, simplified transport models, and does not consider the Soret effect. As a result, it is not clear if the discrepancy results from the modeling approach chosen in \cite{Gaucherand2023} or if they persist when applying detailed methods. \citet{DAlessio2024} computed dispersion relations at an elevated temperature of $500~\mathrm{K}$ and a mixture with $\XHF=0.5$ and $\phi=0.5$ using detailed transport. They investigated the effect of pressure increase from $1~\mathrm{bar}$ to $10~\mathrm{bar}$, showing an increase of instabilities. Furthermore, they analyzed the effect of including the Soret effect in transport modeling, confirming its importance also in \ac{NH3}/\ac{H2} blends.

The present study aims to deliver a comprehensive investigation of dispersion relations for \ac{NH3}/\ac{H2}/air flames, considering a wide range of fuel blend compositions, equivalence ratios, unburned temperatures, and pressures. Specifically, we will answer the following questions:
\begin{itemize}
	\item Can the non-monotonic behavior of \ac{IFI} with respect to $\XHF$ be observed trough dispersion relations, when applying detailed chemistry and transport models? If so, how can the non-monotonic behavior be explained?
	\item What is the influence of increasing pressure at ambient and elevated temperature? These conditions are particularly interesting for storage safety and practical applications.
	\item How are instabilities influenced by equivalence ratio variations at high-pressure-high-temperature conditions?
\end{itemize}

The paper is structured as follows: In \cref{sec:TheoreticalBackground}, the theoretical background for the linear stability analysis is reviewed along with existing models available in the literature. Thereafter, the configuration and numerical setup is presented in \cref{sec:ConfigurationAndNumericalMethods}. The results are subsequently presented and discussed in \cref{sec:ResultsAndDiscussion} with respect to trends, correlations with non-dimensional groups, and comparison to theoretical models. The paper closes with conclusions in \cref{sec:Conclusions}.

\section{Theoretical background}
\label{sec:TheoreticalBackground}

Dispersion relations represent an efficient way to study the impact of instabilities at different conditions. The analysis of weakly perturbed flames does not introduce additional assumptions as the determination of growth rates is well defined. Furthermore, the dispersion relations computed from planar flames can be directly compared to theoretical predictions as the flames are only weakly stretched, which is a typical assumption in theoretical derivations. In the following, existing models used for the comparison with numerical results in this work are summarized.

Following the principals of asymptotic flame theory, several models for dispersion relations have been derived for two-reactant systems using one-step reaction kinetics. Matalon et al.~\cite{Matalon2003} derived a dispersion relation for Lewis numbers close to unity, given by
\begin{equation}
	\overline{\omega}=\omega_0\overline{k}\underbrace{-\delta\left[B_1+\mathit{Ze}\left(\mathit{Le}_{\rm eff}-1\right)B_2+PrB_3\right]}_{\omega_2}\overline{k}^2\,,
	\label{eq:Matalon}
\end{equation}
where $\overline{\omega}=\omega\tau_{\rm F}$ is the normalized growth rate, and $\overline{k}=kl_{\rm F}$ the the normalized wavenumber. Here, the flame time is defined by

\begin{equation}
	\tau_{\rm F}=l_{\rm F}/s_{\rm L}
\end{equation}
\noindent
with the thermal flame thickness $l_{\rm F}$ defined by

\begin{equation}
	l_{\rm F}=\frac{T_{\rm b}-T_{\rm u}}{\max\left(\frac{\rmd T}{\rmd x}\right)}
\end{equation}
\noindent
and the laminar unstretched burning velocity $s_{\rm L}$. The temperatures $T_{\rm u}$ and $T_{\rm b}$ are the temperatures of the unburned and burned gas, respectively.
The first order term in \cref{eq:Matalon} is related to hydrodynamic or Darrieus-Landau effects~\cite{Darrieus1938, Landau1944} with the growth rate

\begin{equation}
	\omega_0=\frac{\sqrt{\sigma^3+\sigma^2-\sigma}-\sigma}{\sigma+1}\,,
	\label{eq:omega_0}
\end{equation}

\noindent 
where $\sigma=\rho_{\rm u}/\rho_{\rm b}$ is the expansion ratio based on the unburned and burned gas densities, $\rho_{\rm u}$ and $\rho_{\rm b}$, respectively. Since $\rho_{\rm u} \geq \rho_{\rm b}$ in all premixed flames, $\omega_0 \geq 0$ so that the term is destabilizing for all wavenumbers.

The second order term in \cref{eq:Matalon}, hereafter referred to as $\omega_2$, describes the influence of thermal, mass, and viscous diffusion. The coefficients $B_1$, $B_2$, and $B_3$ are only functions of $\sigma$ and the temperature dependency of transport coefficients with $B_1\geq1$, $B_2\geq1/2$, and $B_3\geq0$. In this work, the formulation for arbitrary temperature dependencies  for the coefficients $B_i$ is implemented, as presented by Altantzis et al.~\cite{Altantzis2012}. Details are given in Section 1 of the supplementary material. Further coefficients are the ratio between diffusive and thermal flame thicknesses, $\delta=l_{\rm F}/l_{\rm D}$, with $l_{\rm D}=\lambda/(\rho c_{\rm p} s_{\rm L})$, thermal conductivity $\lambda$, specific heat capacity $c_{\rm p}$, Zeldovich number $\mathit{Ze}$, effective Lewis number $\mathit{Le}_{\rm eff}$, and Prandtl number $Pr$.

The Zeldovich number is defined as
\begin{equation}
	\mathit{Ze}=\frac{E}{R}\frac{T_{\rm b}-T_{\rm u}}{T_{\rm b}^2}\:,
	\label{eq:ZeldovichBase}
\end{equation}

\noindent
where $E$ is the activation energy and $R$ is the universal gas constant. Through dimensional analysis, it is found that $\left(\rho_{\rm u} s_{\rm L}\right)^2 \sim \lambda/c_{\rm p} \exp\left(-\frac{E}{RT}\right)$~\cite{Law2000}, so that the activation energy can be determined as~\cite{Law2000}
\begin{equation}
	\frac{E}{R}=-2\frac{\mathrm{d}\left(\ln\left(\rho_{\rm u} s_{\rm L}\right)\right)}{\mathrm{d}(1/T_{\rm b})}\:.
	\label{eq:ActivationEnergy}
\end{equation}

\noindent
The sensitivity of the burning flux on the burned temperature shown in the right-hand-side term of \cref{eq:ActivationEnergy} is determined by a variation of the inert gas fraction $Y_{\ce{N2}}$ by $\pm 0.3\%$~\cite{Sun1999}. It should be noted that the numerical method to determine the differential in \cref{eq:ActivationEnergy}, i.e., a variation of the inert gas fraction~\cite{Sun1999} or through a variation of $T_{\rm b}$ via a variation of $\Tu$~\cite{Kumar2007}, can influence the magnitude of $\mathit{Ze}$. However, the general trends with respect to \cref{eq:Matalon} remain unchanged, as will be shown in \cref{subsec:ComparisonTheoreticalModels}.

The effective Lewis number $\mathit{Le}_{\rm eff}$ was derived by Joulin and Mitani~\cite{Joulin1981} for a two-reactant mixture with excessive reactant $\rm E$ and deficient reactant $\rm D$ with Lewis numbers $\mathit{Le}_{\rm E}$ and $\mathit{Le}_{\rm D}$, respectively, as
\begin{equation}
	\mathit{Le}_{\rm eff}=1+\frac{\left(\mathit{Le}_{\rm E}-1\right)+\left(\mathit{Le}_{\rm D}-1\right)\mathcal{A}}{1+\mathcal{A}}\:.
\end{equation}
\noindent
Here, $\mathcal{A}$ represents the strength of the mixture defined by
\begin{equation}
	\mathcal{A}=1+\mathit{Ze}\left(\varphi-1\right)
\end{equation}
\noindent
with $\varphi=\phi^{-1}$ for lean mixtures and $\varphi=\phi$ for rich mixtures. For the three-reactant mixtures of \ac{NH3}, \ac{H2}, and \ac{O2} considered in this study, the fuel mixture is treated as a unified component with a Lewis number $\mathit{Le}_{\rm Fuel}$. According to Dinkelacker et al.~\cite{Dinkelacker2011}, $\mathit{Le}_{\rm Fuel}$ can be calculated as the sum of the diffusivity-weighted fuel species Lewis numbers $\mathit{Le}_{\rm H_2}$ and $\mathit{Le}_{\rm NH_3}$,

\begin{equation}
	\frac{1}{\mathit{Le}_{\rm Fuel}}=\frac{\XHF}{\mathit{Le}_{\rm H_2}}+\frac{1-X_{\rm H_2,F}}{\mathit{Le}_{\rm NH_3}}\:.
	\label{eq:Le_D}
\end{equation}
\noindent
Other formulations for fuel mixtures, such as a volumetric average~\cite{Muppala2009} or a heat-release-based average~\cite{Law2005} of fuel component Lewis numbers have also been tested. However, as also shown by \citet{Zitouni2023}, the formulation given by \cref{eq:Le_D} exhibits the best agreement, especially for lean mixtures.

It should be noted that one can derive a critical Lewis number $\mathit{Le}_{\rm c}$ from \cref{eq:Matalon} through $\omega_2=0$. This number represents the threshold below which thermo-diffusive processes become destabilizing and is given by
\begin{equation}
	\mathit{Le}_{\rm c} = 1 - \frac{B_1 + \mathit{Pr}B_3}{\mathit{Ze}B_2}\,. \label{eq:Le_c}
\end{equation}

Besides the model by \citet{Matalon2003}, Sivashinsky~\cite{Sivashinsky1977, Sivashinsky1977b} derived an implicit formulation for the dispersion relation given by 

\begin{equation}
	0=\frac{(\mathit{Le}_{\rm eff}-q)(p-r)}{\mathit{Le}_{\rm eff}-q+p-1}-\frac{\mathit{Ze}}{2}\,,
	\label{eq:Sivashinsky}
\end{equation}

\noindent
with
\begin{eqnarray}
	p&=&\frac{1}{2}\left[1+\sqrt{1+4\left(\delta\overline{\omega}+\delta^2\overline{k}^2\right)}\right]	\\
	q&=&\frac{\mathit{Le}_{\rm eff}}{2}\left[1+\sqrt{1+\frac{4\left(\delta\overline{\omega}\mathit{Le}_{\rm eff}+\delta^2\overline{k}^2\right)}{\mathit{Le}_{\rm eff}^2}}\right]	\\
	r&=&\frac{1}{2}\left[1-\sqrt{1+4\left(\delta\overline{\omega}+\delta^2\overline{k}^2\right)}\right]\,.
\end{eqnarray}

\noindent
This formulation is derived for the asymptotic limit of small density variations within the flame, i.e., $\sigma\rightarrow1$, and hence neglects the influence of \acp{DLI}. However, in contrast to \cref{eq:Matalon}, it allows for Lewis numbers sufficiently smaller than unity.
\section{Numerical methods and configuration}
\label{sec:ConfigurationAndNumericalMethods}

In \cref{subsec:NumericalMethodsAndModels}, the applied numerical methods and employed models are introduced. Subsequently, the simulation configuration and the methodology of calculating the dispersion relations is outlined in \cref{subsec:ConfigurationAndDeterminationOfDispersionRelations}.

\subsection{Numerical methods and models}
\label{subsec:NumericalMethodsAndModels}

The \acp{DNS} in this study are performed using PeleLMeX~\cite{PeleLMeX_JOSS, PeleSoftware}. PeleLMeX solves the multi-species reactive Navier-Stokes equations in the low-Mach formulation~\cite{day2000numerical}. The equations are advanced in time through a spectral-deferred corrections approach that conserves species, mass and energy~\cite{nonaka2012deferred, nonaka2018conservative}. The advection term is discretized with a second order Godunov scheme. The energy and species equations are treated implicitly utilizing the ODE solver CVODE from the SUNDIALS package~\cite{hindmarsh2005sundials}. PeleLMeX features (adaptive) mesh refinement inherited from the AMReX package~\cite{AMReX_JOSS}. Note that within the scope of this work, a static mesh refinement is applied, since otherwise the refinement criteria would need to be different for each case.

Chemical reactions and associated rates are modeled through the reaction mechanism developed by Zhang et al.~\cite{Zhang2021} (30 species, 243 reactions). This model demonstrated very good overall performance in terms of quantitative agreement with comprehensive experimental data for flame speed, ignition delay time, and species concentrations~\cite{Girhe2024}. Assumptions related to the reduction of chemical mechanisms are avoided by using detailed chemistry. Transport is modeled with a mixture-average approach, where the individual species viscosity $\mu_i$, conductivity $\lambda_i$, and binary diffusion coefficients $D_{i,j}$ are obtained from pre-computed logarithmic temperature fits~\cite{Ern1994}. More specifically, $\mu_i$ and $D_{i,j}$ are calculated via kinetic theory, while taking the polar-non-polar interaction into account~\cite{Hirschfelder1964} when interpolating for the tabulated collision integrals~\cite{Monchick1961}. For $\lambda_i$, the formulation by \citet{Warnatz1982} is applied. The viscosity and conductivity of the mixture are evaluated from the polynomial formulation in~\cite{Ern1995}. The Soret effect is included as presented in~\cite{Howarth2024}.

\subsection{Configuration and determination of dispersion relations}
\label{subsec:ConfigurationAndDeterminationOfDispersionRelations}

In this study, simulations of two-dimensional, planar laminar premixed flames with air as oxidizer and an \ac{NH3}/\ac{H2} blend as fuel are conducted. The simulation domain is a rectangular box with periodic boundary conditions in cross-wise ($x$) direction and inflow and outflow conditions in stream-wise ($y$) direction. The dimensions of the isotropic base grid are $L_x \times L_y = 80\lF \times 20\lF$ with a base resolution of $\Delta x_{\rm Base} = \lF / 12.8$. Two levels of static mesh refinement are applied in the vicinity of the flame, i.e., between $y=\left[7\lF, 13\lF\right]$, to resolve the flame with $\Delta x_{\rm Fine} = \lF / 51.2$. Increasing the resolution further, either by running with more levels of mesh refinement or a simulation without mesh refinement with equivalent resolution, does not alter the results. More details are provided in Section 2 of the supplementary material. The domain is initialized by mapping a one-dimensional flamelet solution obtained by FlameMaster~\cite{Pitsch1998} onto the domain along the $y$-direction. Thereby the flame, defined through the steepest temperature gradient, is located at $y_0=10\lF$. To apply an initial perturbation, the method developed by \citet{AlKassar2024} is applied. Here, the initial solution is perturbed by a superposition of $N$ harmonic functions with wavenumbers  $k_i= 2\pi i /L_x$,
\begin{equation}
y_{\rm Flame}(x, t=0) = y_0 + A_0 \sum_{i=1}^{N} \sin\left(i \frac{2\pi}{L_x} x + \psi(i) \right)\,.
\end{equation}
For this study, the initial perturbation amplitude is $A_0=10^{-6}\lF$, and the total number of wavenumbers is $N=40$. This results in a wavelength range of $\lambda\in[2\lF, 80\lF]$, and normalized wavenumber range of $\overline{k}\in[\pi/40, \pi]$. The phase shift $\psi(i)$ is a random value between $[0, 2\pi]$. Note that the same random seed is used for all simulations to ensure reproducibility. The advantage of the method by \citet{AlKassar2024} is that a full dispersion relation can be extracted from only one simulation, since the growth rates $\omega(k_i)$ are independent in the linear regime. For further details on the method as well as an extensive validation, the reader is referred to \cite{AlKassar2024}.

\begin{figure}[h!]
	\centering
	\ifdefined\preprint
	\begin{subfigure}[b]{0.49\linewidth}
	\fi
	\ifdefined\final
	\begin{subfigure}[b]{\linewidth}
	\fi
		\centering
		\includegraphics[width=\linewidth]{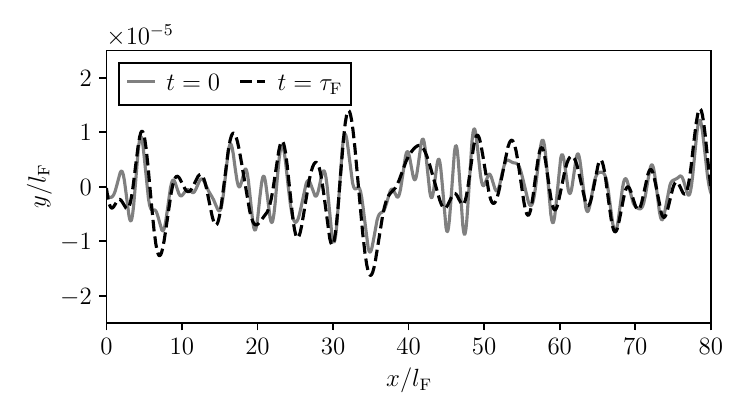}
		\caption{$\XHF=1.0$}
		\label{fig:TemporalGrowthRate:FlameFront}
	\end{subfigure}
	\ifdefined\preprint
	\begin{subfigure}[b]{0.24\linewidth}
	\fi
	\ifdefined\final
	\begin{subfigure}[b]{0.49\linewidth}
	\fi
		\centering
		\includegraphics[width=\linewidth]{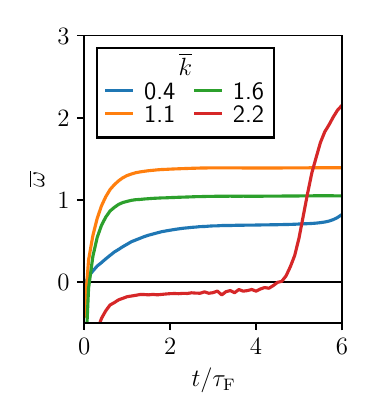}
		\caption{$\XHF=0.4$}
		\label{fig:TemporalGrowthRate:TemporalGrowthRate}
	\end{subfigure}
	\ifdefined\preprint
	\begin{subfigure}[b]{0.24\linewidth}
	\fi
	\ifdefined\final
	\begin{subfigure}[b]{0.49\linewidth}
	\fi
		\centering
		\includegraphics[width=\linewidth]{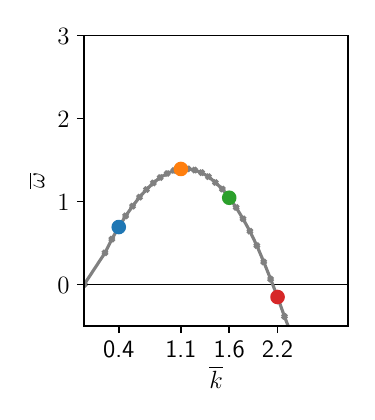}
		\caption{$\XHF=0.4$}
		\label{fig:TemporalGrowthRate:DispRelConstruction}
	\end{subfigure}

	\caption{Procedure for the numerical determination of linear growth rates for an example case ($\phi = 0.5$, $\XHF = 1.0$, $p = 1~\rm bar$, $\Tu = 298~\rm K$): (a) Temporal evolution of the weakly perturbed flame front. (b) Temporal growth rate $\omega(t,k)$ for different wavenumbers $k$. After an initialization phase, a plateau is observed, from where the linear growth rate $\omega(k)$ is determined. (c) Extracted $\overline{\omega}$ plotted over $\overline{k}$ yields the dispersion relation for the simulated case.}
	\label{fig:TemporalGrowthRate}
\end{figure}

Over the course of the simulation, the flame surface $y_{\rm Flame}(x, t)$ is traced in space and time through an iso-surface at $T=1000~\rm K$. The specific selection of variable and the iso-value does not influence the results, as shown in Section 3 of the supplementary material. Using a Fourier transformation, the temporal evolution of the frequency-specific amplitude $A(t,k_i)$ is extracted. The temporal growth rate $\omega(t,k_i)$ is then determined as 
\begin{equation}
	\omega(t,k_i)=\frac{\rmd \ln\left(A(t,k_i)\right)}{\rmd t}\,.
\end{equation}
\noindent
\Cref{fig:TemporalGrowthRate} shows $\omega(t,k_i)$ for different $k_i$. The linear growth rate $\omega(k_{i})$ is determined in the temporal plateau for each wavelength $k_{i}$, as seen in \cref{fig:TemporalGrowthRate}.

\begin{figure}[h]
	\centering
	\ifdefined\preprint
	\includegraphics[width=0.5\linewidth]{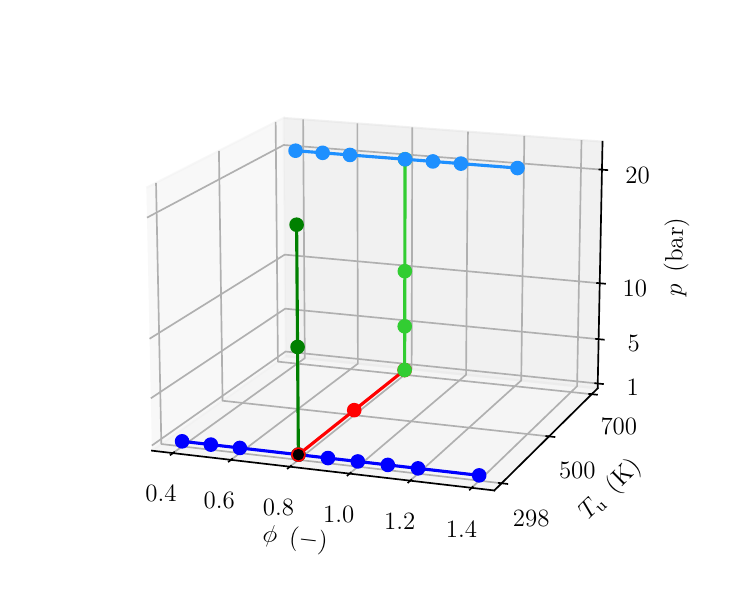}
	\fi
	\ifdefined\final
	\includegraphics[width=\linewidth]{ParamSpace.pdf}
	\fi
	\caption{Visualization of the systematical screening of the parameter space. Each point represents up to 6 different values of $\XHF$.}
	\label{fig:ParamSpace}
\end{figure}

This study features a wide range of parameters covering typical conditions in applications such as industrial furnaces and gas turbines. The equivalence ratio $\phi$ is varied between lean and rich with $\phi\in\{0.4, 0.5, 0.6, 0.8, 0.9, 1.0, 1.1, \allowbreak 1.2, 1.4\}$. The temperature in the unburned is $T_{\rm u}\in\{298~\mathrm{K}, 500~\mathrm{K}, 700~\mathrm{K}\}$ and the pressure is $p\in\{1~\mathrm{bar},5~\mathrm{bar},10~\mathrm{bar},20~\mathrm{bar}\}$. For a systematic analysis, the parameter space is screened through five variations detailed in \cref{tab:variations} and visualized in \cref{fig:ParamSpace}.
\begin{table}[h]
	\centering
	\caption{Systematic parameter sweeps conducted in this study.}
	\begin{tabular}{c c c c}
		Variation & $\phi$ & $\Tu\rm~(K)$ & $p\rm~(bar)$\\
		\hline
		V1 & $0.4$ - $1.4$ & $298$ & $1$\\
		V2 & $0.8$ & $298$ & $1, 10, 20$\\
		V3 & $0.8$ & $298, 500, 700$ & $1$\\
		V4 & $0.8$ & $700$ & $1, 5, 10, 20$\\
		V5 & $0.6$ - $1.1$ & $700$ & $20$
	\end{tabular}
	\label{tab:variations}
\end{table}

The variations V1, V2, and V3 examine the effect of changing $\phi$, $p$, and $\Tu$, respectively, starting from a reference case at $\phi = 0.8$, $p = 1~\rm bar$, and $\Tu = 298~\rm K$. Additionally, a variation of $p$ at high $\Tu$ is conducted in V4, and a variation of $\phi$ at high $p$ and $\Tu$ is conducted in V5. For each condition, the \ac{H2} fraction in the fuel blend $\XHF$ is varied between pure \ac{NH3} ($\XHF=0.0$) and pure \ac{H2} ($\XHF=1.0$). It is worth noting that pure \ac{NH3}/air flames at ambient conditions exhibit a lower flammability limit of $\phi_{\rm FL,low}\approx0.7$ and no flame can be observed below this equivalence ratio. Hence, these conditions are not included in this study. Additional cases are included for $\phi=0.5$, $\XHF=0.5$, and $\Tu=500~\rm K$ at $p=1~\rm bar$ and $p=10~\rm bar$ for direct comparison with the studies of \citet{DAlessio2024} and \citet{Gaucherand2023}. In total, 130 different conditions are analyzed. A complete list of cases is provided in Section 4 of the supplementary material.

\section{Results and discussion}
\label{sec:ResultsAndDiscussion}

In \cref{subsec:ComparisonToOtherStudies}, the computed dispersion relations are firstly compared to existing studies. In \cref{subsec:DiscussionDispersionRelations}, the effects of \ac{H2} fraction, equivalence ratio, pressure, and unburned temperature are discussed. In \cref{subsec:PressureZeldovich}, an investigation of the pressure dependency of the Zeldovich number is presented. Finally, the numerically obtained dispersion relations are compared to theoretical models in \cref{subsec:ComparisonTheoreticalModels}.

\subsection{Comparison to other studies}
\label{subsec:ComparisonToOtherStudies}

\begin{figure}[h!]
	\centering
	\ifdefined\preprint
	\begin{subfigure}[b]{0.49\linewidth}
	\fi
	\ifdefined\final
	\begin{subfigure}[b]{\linewidth}
	\fi
		\centering
		\includegraphics[width=\linewidth]{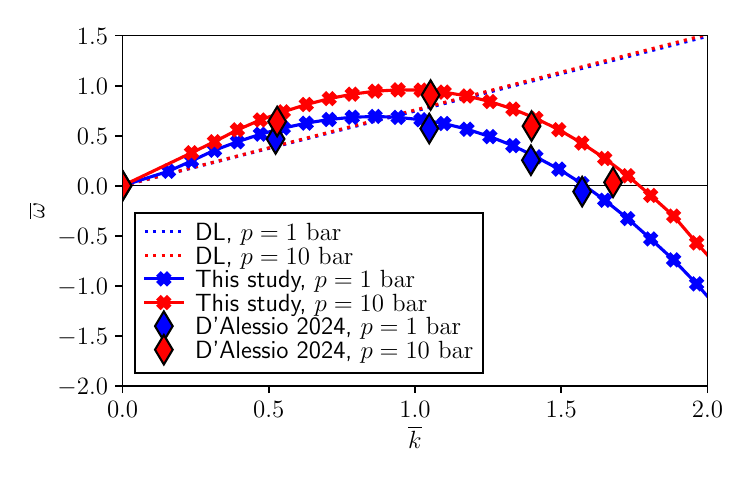}
		\caption{}
		\label{fig:Comparison:DAlessio}
	\end{subfigure}
	\ifdefined\preprint
	\begin{subfigure}[b]{0.49\linewidth}
	\fi
	\ifdefined\final
	\begin{subfigure}[b]{\linewidth}
		\fi
		\centering
		\includegraphics[width=\linewidth]{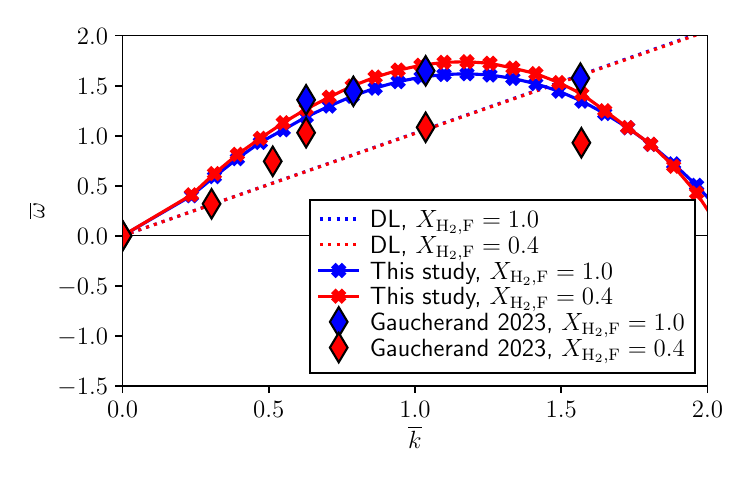}
		\caption{}
		\label{fig:Comparison:Gaucherand}
	\end{subfigure}
	\caption{Comparison of dispersion relations with results from other studies. (a): Comparison with \cite{DAlessio2024} at $p=1\rm~bar$ (blue) and $p=10\rm~bar$ (red) for $\XHF=0.5$, $\phi=0.5$, $\Tu=500\rm~K$. (b): Comparison with \cite{Gaucherand2023} for blends with $\XHF=1.0$ (blue) and $\XHF=0.4$ (red) for $\phi=0.4$, $\Tu=298\rm~K$, and $p=1\rm~bar$. In both figures, dashed lines represent the Darrieus-Landau growth rate, cross symbols represent the results from this study, and diamond symbols represent the results from the comparison study.}
	\label{fig:Comparison}
\end{figure}

\Cref{fig:Comparison} shows a comparison of results from the present study with data from \citet{DAlessio2024} and \citet{Gaucherand2023}. \Cref{fig:Comparison:DAlessio} shows good agreement between the data sets from \citet{DAlessio2024} and the present study with only minor differences. As both studies employ similar transport models, i.e., a mixture average model with direct calculation of species transport properties, these differences are most likely related to different numerical methods or the specific normalization. Deviations in \cref{fig:Comparison:Gaucherand} are more prominent, exhibiting an inversion of the instability trend in terms of the impact of the hydrogen fraction. These differences may stem from the simplified transport models or reduced chemistry in \citet{Gaucherand2023}. However, identifying the exact origins is beyond the scope of this work.

\subsection{Numerically computed dispersion relations}
\label{subsec:DiscussionDispersionRelations}

\begin{figure}
	\centering
	\ifdefined\preprint
	\begin{subfigure}[b]{0.49\linewidth}
	\fi
	\ifdefined\final
	\begin{subfigure}[b]{\linewidth}
	\fi
		\centering
		\includegraphics[width=\linewidth]{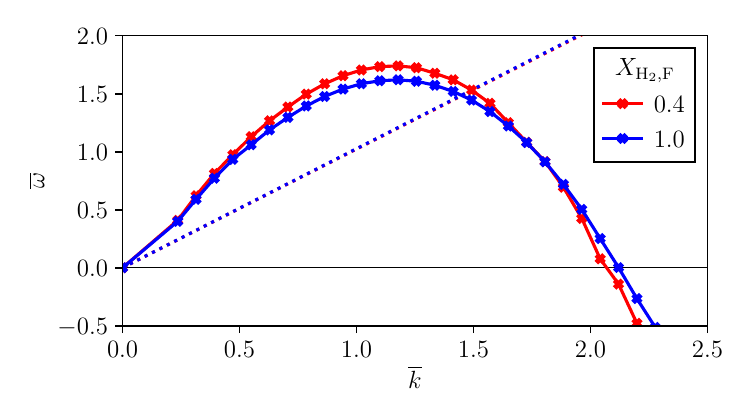}
		\caption{$\phi=0.4$}
		\label{fig:DispRel04}
	\end{subfigure}
	\ifdefined\preprint
	\begin{subfigure}[b]{0.49\linewidth}
	\fi
	\ifdefined\final
	\begin{subfigure}[b]{\linewidth}
	\fi
		\centering
		\includegraphics[width=\linewidth]{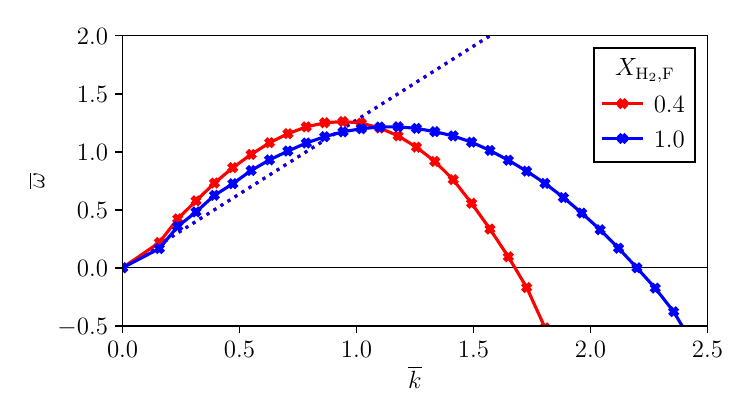}
		\caption{$\phi=0.6$}
		\label{fig:DispRel06}
	\end{subfigure}
	\ifdefined\preprint
	\begin{subfigure}[b]{0.49\linewidth}
	\fi
	\ifdefined\final
	\begin{subfigure}[b]{\linewidth}
		\fi
		\centering
		\includegraphics[width=\linewidth]{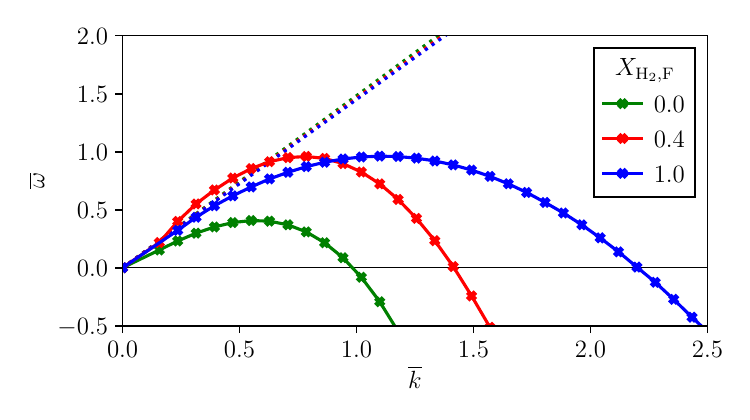}
		\caption{$\phi=0.8$}
		\label{fig:DispRel08}
	\end{subfigure}
	\ifdefined\preprint
	\begin{subfigure}[b]{0.49\linewidth}
	\fi
	\ifdefined\final
	\begin{subfigure}[b]{\linewidth}
		\fi
		\centering
		\includegraphics[width=\linewidth]{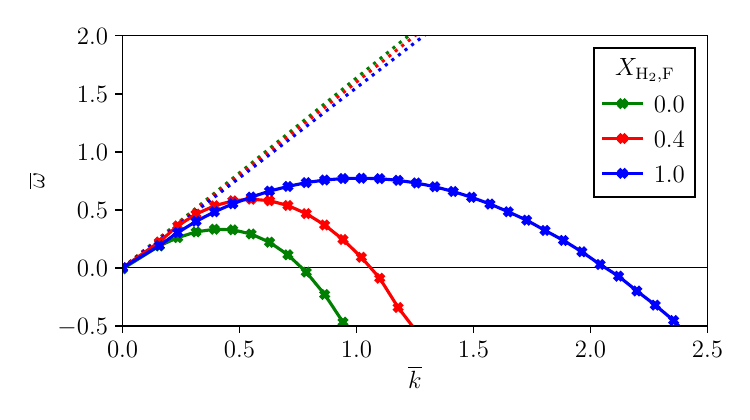}
		\caption{$\phi=1.0$}
		\label{fig:DispRel10}
	\end{subfigure}
	\caption{Dispersion relations at four different equivalence ratios for pure \ce{NH3} (green), pure \ce{H2} (blue), and an \ce{NH3}/\ce{H2} blend with $\XHF=0.4$ (red). Symbols represent the numerically determined growth rates connected through solid lines, and dotted lines represent the theoretical growth rate of hydrodynamic instabilities, see \cref{eq:omega_0}. All simulations are conducted at ambient conditions, $T_{\rm u}=298~\rm K$ and $p=1~\rm bar$.}
	\label{fig:DispRel_phi}
\end{figure}

\begin{figure}[h!]
	\centering
	\ifdefined\preprint
	\begin{subfigure}[b]{0.49\linewidth}
	\fi
	\ifdefined\final
	\begin{subfigure}[b]{\linewidth}
	\fi
		\centering
		\includegraphics[width=\linewidth]{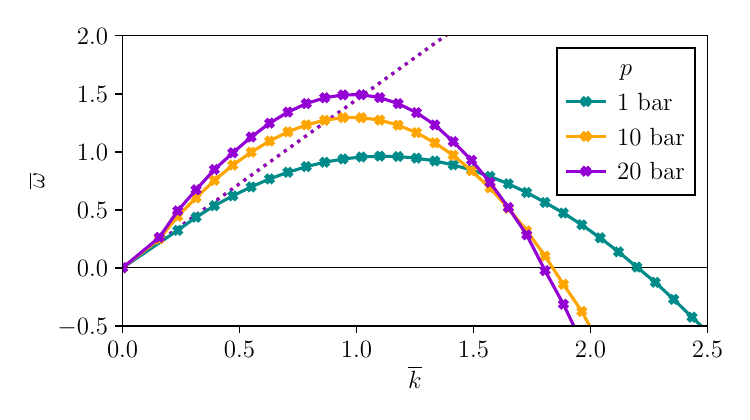}
		\caption{$\XHF=1.0$}
		\label{fig:DispRelp100}
	\end{subfigure}
	\ifdefined\preprint
	\begin{subfigure}[b]{0.49\linewidth}
	\fi
	\ifdefined\final
	\begin{subfigure}[b]{\linewidth}
	\fi
		\centering
		\includegraphics[width=\linewidth]{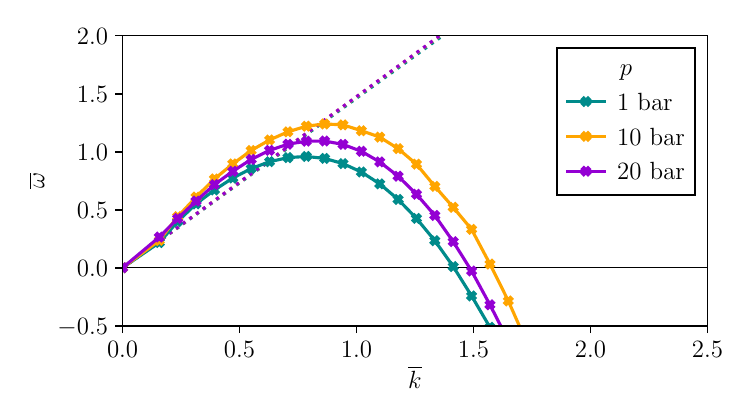}
		\caption{$\XHF=0.4$}
		\label{fig:DispRelp040}
	\end{subfigure}
	\caption{Dispersion relations at different pressures, i.e., $p=1~\rm bar$ (cyan), $p=10~\rm bar$ (orange), and $p=20~\rm bar$ (violet), (a) for pure \ce{H2}, and (b) an \ce{NH3}/\ce{H2} blend with $\XHF=0.4$. Symbols represent the numerically determined growth rates connected through solid lines, and dotted lines represent the theoretical growth rate of hydrodynamic instabilities, see \cref{eq:omega_0}. All simulations are conducted at ambient conditions, $T_{\rm u}=298~\rm K$ and $\phi=0.8$.}
	\label{fig:DispRel_pressure}
\end{figure}

\begin{figure*}
	\centering
	\begin{subfigure}[b]{\textwidth}
		\centering
		\includegraphics[width=\textwidth]{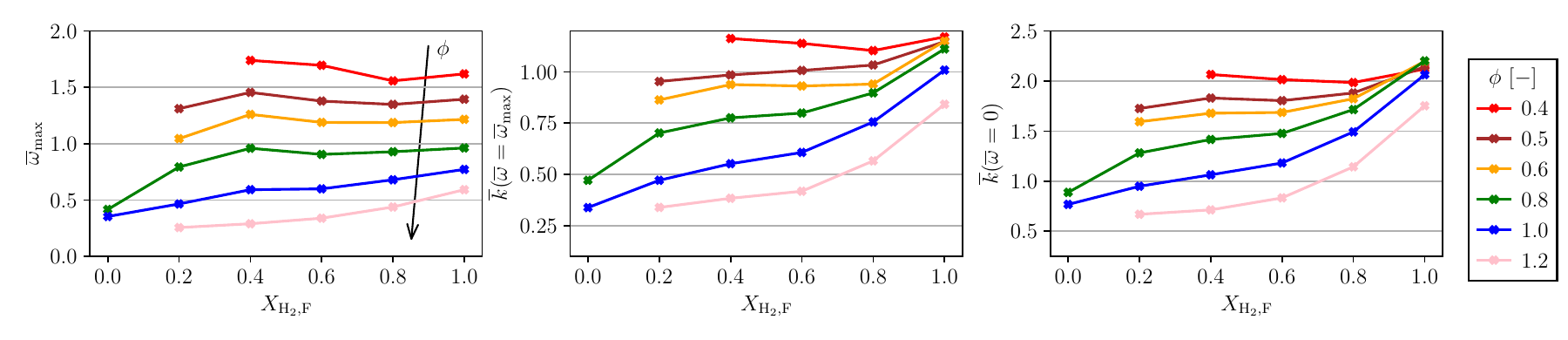}
		\caption{V1: Variation of $\phi$ and $\XHF$ at constant $p=1\rm~bar$ and $\Tu=298\rm~K$.}
		\label{fig:Derived_V1}
	\end{subfigure}
	\begin{subfigure}[b]{\textwidth}
		\centering
		\includegraphics[width=\textwidth]{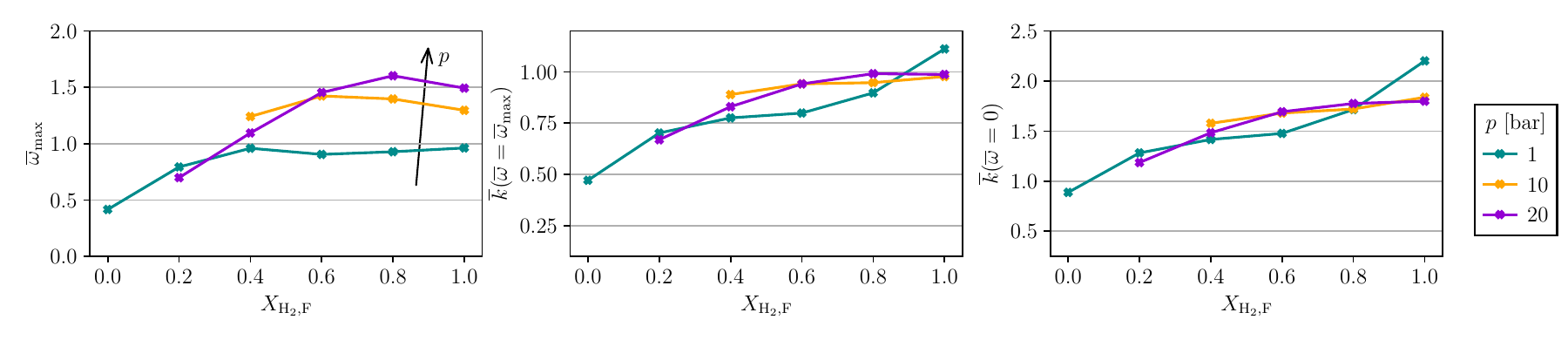}
		\caption{V2: Variation of $p$ and $\XHF$ at constant $\phi=0.8$ and $\Tu=298\rm~K$.}
		\label{fig:Derived_V2}
	\end{subfigure}
	\begin{subfigure}[b]{\textwidth}
		\centering
		\includegraphics[width=\textwidth]{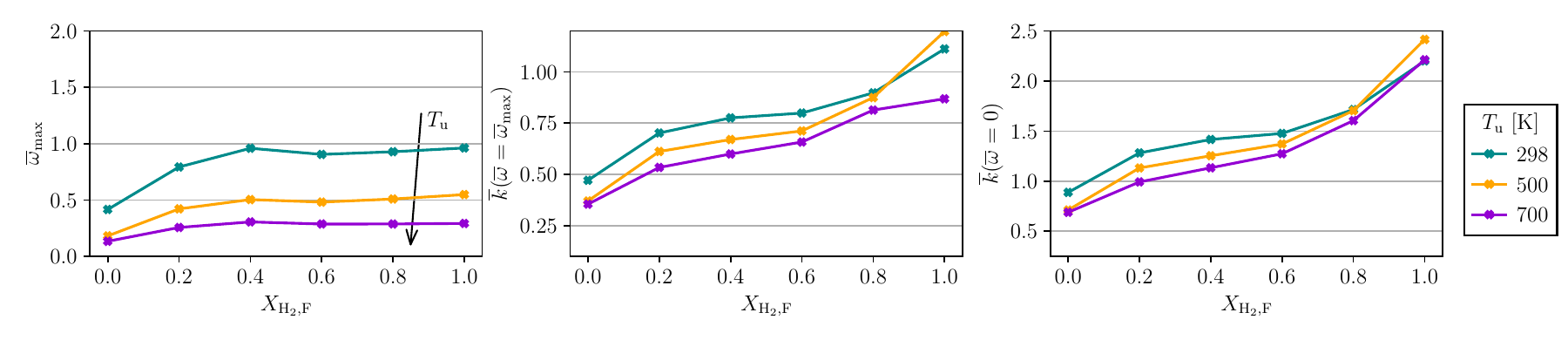}
		\caption{V3: Variation of $\Tu$ and $\XHF$ at constant $\phi=0.8$ and $p=1\rm~bar$.}
		\label{fig:Derived_V3}
	\end{subfigure}
	\begin{subfigure}[b]{\textwidth}
		\centering
		\includegraphics[width=\textwidth]{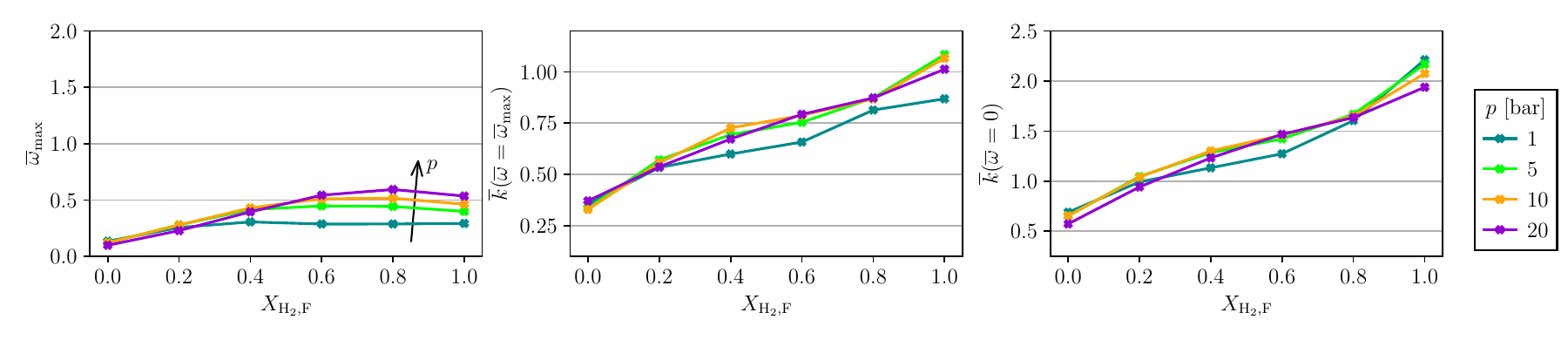}
		\caption{V4: Variation of $p$ and $\XHF$ at constant $\phi=0.8$ and $\Tu=700\rm~K$.}
		\label{fig:Derived_V4}
	\end{subfigure}
	\begin{subfigure}[b]{\textwidth}
		\centering
		\includegraphics[width=\textwidth]{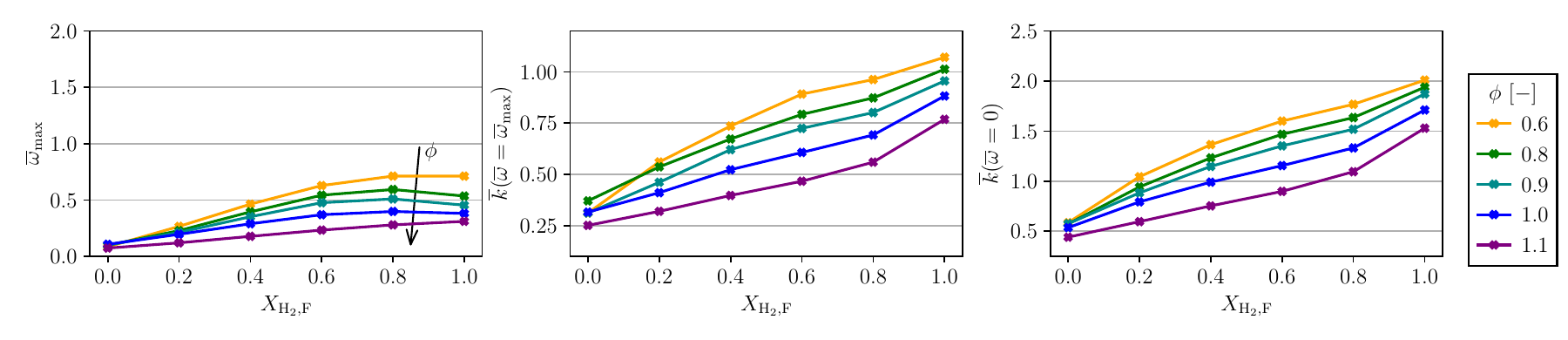}
		\caption{V5: Variation of $\phi$ and $\XHF$ at constant $p=20\rm~bar$ and $\Tu=700\rm~K$.}
		\label{fig:Derived_V5}
	\end{subfigure}
	\caption{Peak growth rate $\overline{\omega}_{\rm max}$ (left), wavenumber $\overline{k}(\overline{\omega}=\overline{\omega}_{\rm max})$ (center), and cut-off wave length $\overline{k}(\overline{\omega}=0)$ (right) over $\XHF$ for variations V1 to V5.}
	\label{fig:MaxAndCutoff}
\end{figure*}

\begin{figure*}
	\centering
	\begin{subfigure}[b]{\textwidth}
		\centering
		\includegraphics[width=\textwidth]{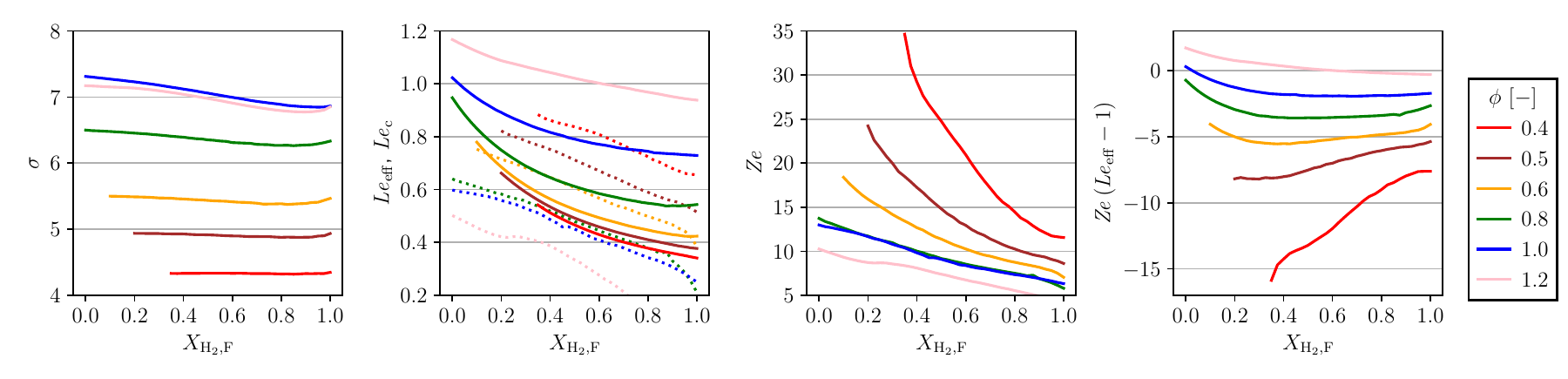}
		\caption{Variation of $\phi$ and $\XHF$ at constant $p=1\rm~bar$ and $\Tu=298\rm~K$.}
		\label{fig:NonDimGroups_phi}
	\end{subfigure}
	\begin{subfigure}[b]{\textwidth}
		\centering
		\includegraphics[width=\textwidth]{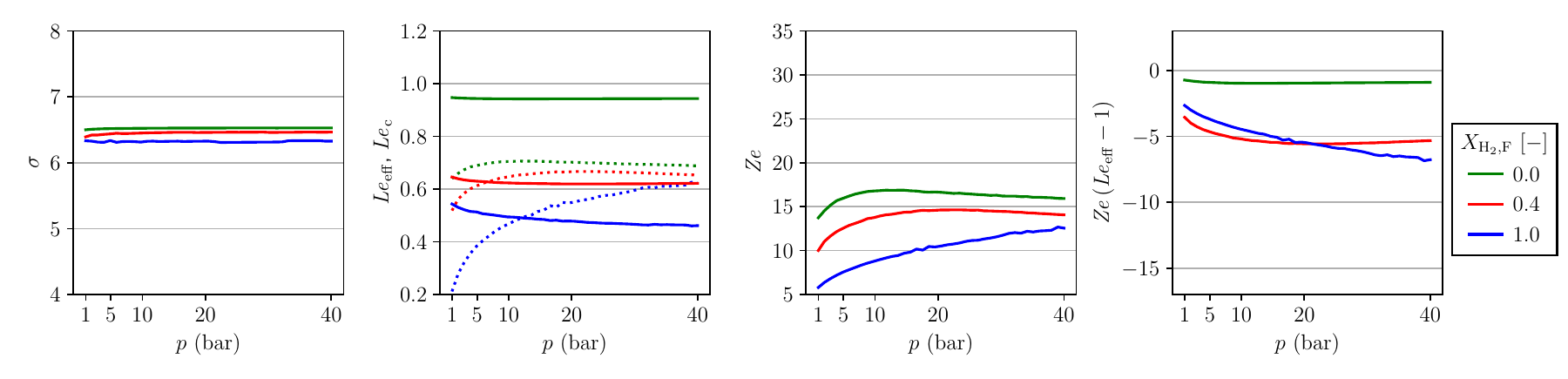}
		\caption{Variation of $p$ and $\XHF$ at constant $\phi=0.8$ and $\Tu=298\rm~K$.}
		\label{fig:NonDimGroups_pressure}
	\end{subfigure}
	\caption{Non-dimensional groups for variations of (a) $\XHF$ and $\phi$, and (b) $p$ and $\XHF$. Expansion ratio $\sigma$ (outer left), effective Lewis number $\mathit{Le}_{\rm eff}$ (solid) and critical Lewis number $\mathit{Le}_{\rm c}$ (dotted) (center left), Zeldovich number $\mathit{Ze}$ (center right), and $\mathit{Ze}\left(\mathit{Le}-1\right)$ (outer right). Note that the $x$-axis for (a) presents $\XHF$ while presenting $p$ in (b).}
\label{fig:NondimensionalGroups}
\end{figure*}

In the following sections, the effect of variations of $\phi$, $p$, and $\Tu$ will be discussed based on the numerically computed dispersion relations (\cref{fig:DispRel_phi,,fig:DispRel_pressure}), their characteristic features, i.e.,~the peak growth rate $\overline{\omega}_{\rm max}$, wavenumber $\overline{k}(\overline{\omega}=\overline{\omega}_{\rm max})$, and cut-off wavenumber $\overline{k}(\overline{\omega}=0)$ (\cref{fig:MaxAndCutoff}), and selected non-dimensional groups (\cref{fig:NondimensionalGroups}). Note that the full dispersion relations are  depicted only for selected cases in variations V1 and V2 in \cref{fig:DispRel_phi,,fig:DispRel_pressure}. For all other cases, solely the characteristic features are presented in the following. Additionally, a comprehensive collection of all dispersion relations is provided in Section 5 of the supplementary material.

\subsubsection{Variation of $\phi$ and $\XHF$ (V1)}

\Cref{fig:DispRel_phi} shows the numerically determined dispersion relations (solid lines) for variations of $\phi$ (V1), and the corresponding theoretically derived hydrodynamic instability growth rates based on \cref{eq:omega_0} (dashed lines). For lean and stoichiometric mixtures shown in \cref{fig:DispRel_phi}, flames with $\XHF=0.4$ exhibit larger growth rates at low wavenumbers than flames of the pure components. This indicates a non-monotonic effect of \ce{H2}-addition on \acp{IFI}. This effect can also be seen in \cref{fig:Derived_V1}, represented by the non-monotonic behavior of $\overline{\omega}_{\rm max}$ with $\XHF$ for all mixtures with $\phi\leq0.8$. It is worth noting that the local maxima can be found at $\XHF=0.4$ for any lean mixture analyzed within the scope of this work. These findings complement those of \citet{Ichikawa2015} and \citet{Zitouni2023}. For the variation of $\phi$, the strongest instabilities are found for molar \ac{H2} fractions around $40\%$. For rich mixtures, pure \ac{H2}/air flames exhibit stronger \acp{IFI} than blends of \ac{NH3} and \ac{H2} or even pure \ac{NH3}/air flames, and dispersion relations are strictly below the hydrodynamic growth rate, indicating the stabilizing effect of thermo-diffusive processes. See also Section 5 of the supplementary material.

To understand the non-monotonic behavior, it is useful to examine the non-dimensional numbers appearing in \cref{eq:Matalon}, i.e., the expansion ratio $\sigma$, effective Lewis number $\mathit{Le}_{\rm eff}$, and the Zeldovich number $\mathit{Ze}$, as well as their combined effect in $\mathit{Ze}\left(\mathit{Le}-1\right)$, given in \cref{fig:NonDimGroups_phi}. Note that the latter is not an independent non-dimensional group, but is only chosen to represent the joint effect of the Zeldovich and Lewis numbers.

\Cref{fig:NondimensionalGroups} reveals that $\sigma$ is only weakly dependent on $\XHF$. The expansion ratio can also be written as $\sigma=\rho_{\rm u}/\rho_{\rm b}=(M_{\rm u}/M_{\rm b})\cdot(T_{\rm b}/T_{\rm u})$, where $M_{\rm u}$ and $M_{\rm b}$ are the molar masses in the unburned and burned mixture, respectively. While the ratio $T_{\rm b}/T_{\rm u}$ increases for increasing $\XHF$ due to an increasing adiabatic flame temperature, the opposite holds true for $M_{\rm u}/M_{\rm b}$, resulting in only minor changes of the expansion ratio. This leads to only small changes in the hydrodynamic growth rates, which are almost invariant with respect to a variation of the fuel blending ratio. Regarding the magnitude of the peak growth rate, \citet{Berger2022} showed that for pure \ac{H2}/air flames it increases with increasing $\mathit{Ze}$ and decreasing $\mathit{Le}_{\rm eff}$. The same trend can readily be observed for $\phi$ variations for \ac{NH3}/\ac{H2}/air flames at fixed $\XHF$, since Zeldovich number and effective Lewis number show opposing trends for this variation. For a variation of $\XHF$ at given $\phi$, however, $\mathit{Ze}$ and $\mathit{Le}_{\rm eff}$ show the same trend, i.e., a decrease with increasing $\XHF$. The joint effect of both parameters is captured through the term $\mathit{Ze}\left(\mathit{Le}_{\rm eff}-1\right)$, which also appears in $\omega_2$. 
Lower values imply stronger thermo-diffusive instabilities. As expected, $\mathit{Ze}\left(\mathit{Le}_{\rm eff}-1\right)$ decreases with decreasing $\phi$. Furthermore, for lean conditions, the term becomes non-monotonic with respect to a variation of $\XHF$, revealing a minimum at $\XHF=0.4$. The non-monotonic behavior becomes more pronounced as $\phi$ decreases and vanishes for rich mixtures. Hence, this coupling reveals the reason for the non-monotonic trend of \acp{TDI} in \ce{NH3}/\ce{H2}/air flames: at first, the flame becomes weaker with decreasing $\XHF$, represented by the increasing Zeldovich number, and with this, more perceptive for instabilities. As a result, \acp{TDI} first increase with decreasing $\XHF$. However, the effective Lewis number also increases with decreasing $\XHF$,  indicating a reduced imbalance between heat and species diffusion. For the limit of $\XHF\rightarrow0$, $\mathit{Le}_{\rm eff} \rightarrow 1$, so that the \acp{TDI} vanish. This principle can also be visualized through the comparison of $\mathit{Le}_{\rm eff}$ and the critical Lewis number $\mathit{Le}_{\rm c}$ as defined in \cref{eq:Le_c} and shown as dotted lines in \cref{fig:NondimensionalGroups}. Since both numbers are decreasing with increasing $\XHF$, their difference $\mathit{Le}_{\rm eff}-\mathit{Le}_{\rm c}$, and with this their level of instability, shows a non-monotonic behavior.

A similar, non-linear trend as for the peak growth rate can be observed for the normalized wavenumber at maximum growth rate $\overline{k}(\overline{\omega}=\overline{\omega}_{\rm max})$ and the cut-off wavenumber $\overline{k}(\overline{\omega}=0)$, i.e., the largest wavenumber with non-negative growth rate. This may indicate a decrease of the characteristic length scale of fully developed \acp{TDI}~\cite{Berger2019}. However, the proof of generality for this correlation within the context of \ac{NH3}/\ac{H2}/air flames requires the analysis of the non-linear flame evolution and is not part of this study.

\subsubsection{Variation of $p$ at low $\Tu$ (V2)}

The effects of pressure on \acp{IFI}, analyzed through variation V2, are depicted in \cref{fig:DispRel_pressure,,fig:Derived_V2}. For pure \ac{H2}/air flames at conditions considered here, instabilities increase with increasing pressure, as previously reported by \cite{Berger2022}. For \ac{NH3}/\ac{H2} blends with $\XHF=0.4$, however, a non-monotonic behavior is observed, first showing an increase of \acp{IFI} with $p$ up to $10\rm~bar$ followed by a decrease of \acp{IFI} with further increase of $p$. The same trends are also reflected in the $\mathit{Ze}\left(\mathit{Le}_{\rm eff}-1\right)$ term for pure \ac{H2}/air and a blend, as depicted in \cref{fig:NonDimGroups_pressure}. Furthermore, it is apparent that the non-monotonicity is mainly driven by the Zeldovich number, since $\sigma$ and $\mathit{Le}_{\rm eff}$ remain almost constant with variation of $p$. A detailed reasoning linked to chemical kinetics will be explored in \cref{subsec:PressureZeldovich}. Again, similar trends can be observed for $\overline{k}(\overline{\omega}=\overline{\omega}_{\rm max})$ and $\overline{k}(\overline{\omega}=0)$.

\subsubsection{Additional variations of $\Tu$, $p$, and $\phi$ (V3, V4, and V5)}
The influence of the unburned temperature $\Tu$ is analyzed in variation V3 shown in \cref{fig:Derived_V3}. As discussed in the literature~\cite{Berger2022, Howarth2022}, increasing $\Tu$ decreases \acp{IFI} and this effect is observed to hold true regardless of the \ac{NH3}/\ac{H2} blend ratio. This leads to an overall reduction in instability for the $p$ variation at high $\Tu$ (V4, \cref{fig:Derived_V4}), while the non-monotonic behavior for blends with $\XHF\leq0.4$ remains visible, although not as prominent as at lower $\Tu$. Finally, a variation of $\phi$ and blend fraction at high $p$ and high $\Tu$ (V5, \cref{fig:Derived_V5}) combines the previously discussed effects. Peak growth rates are generally smaller compared to their low-$p$ low-$\Tu$ equivalent. Although still notable, the non-monotonic behavior with respect to $\XHF$ is less prominent and the peak is shifted to higher blend ratios at around $\XHF = 0.8$.

For all three variations (V3 - V5), $\overline{k}(\overline{\omega}=\overline{\omega}_{\rm max})$ and $\overline{k}(\overline{\omega}=0)$ increase almost linearly with $\XHF$ and no non-monotonic trend is visible. Furthermore, the influence of $\Tu$ (V3, \cref{fig:Derived_V3}) and especially $p$ at high $Tu$ (V4, \cref{fig:Derived_V4}) is relatively small, compared to the variation of $\phi$ at high $p$ and $Tu$ (V5, \cref{fig:Derived_V5}).

\subsection{Pressure dependency of the Zeldovich number}
\label{subsec:PressureZeldovich}
\begin{figure}[h!]
	\centering
	\ifdefined\preprint
	\includegraphics[width=0.5\linewidth]{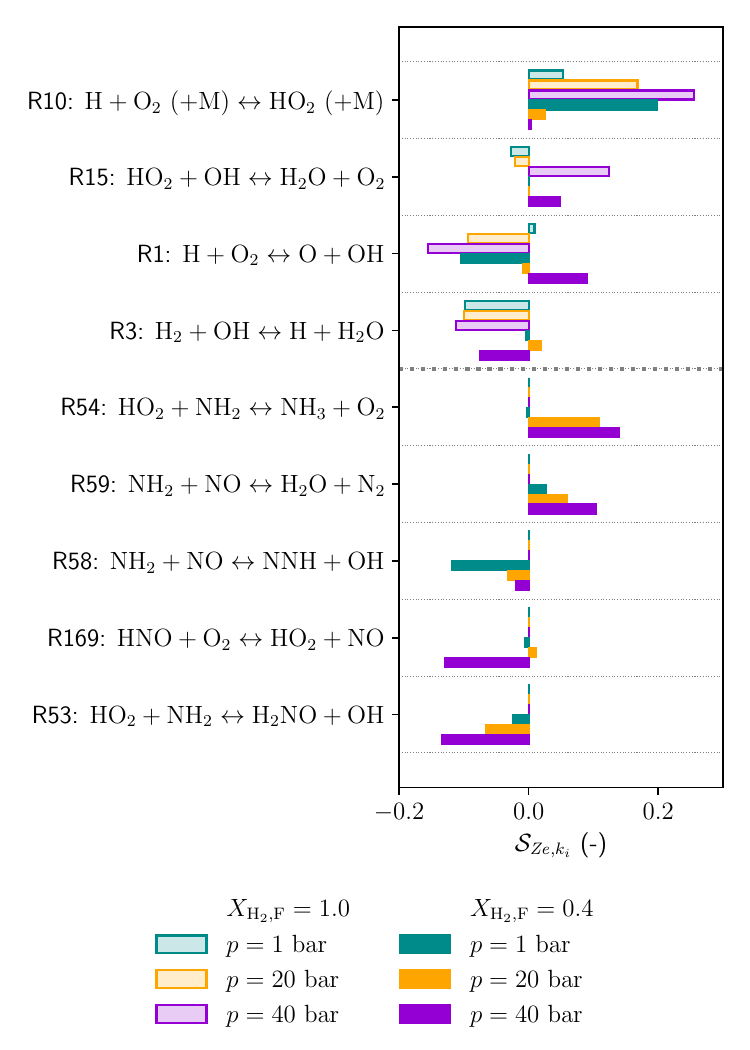}
	\fi
	\ifdefined\final
	\includegraphics[width=\linewidth]{Sensitivities.pdf}
	\fi
	\caption{Selected normalized sensitivity coefficients $\mathcal{S}_{\mathit{Ze},k_i}$ of $\mathit{Ze}$ on the elementary reaction rates $k_i$ for varying pressure. Translucent bars represent pure \ac{H2} as fuel, opaque bars represent an \ac{NH3}/\ac{H2} blend with $\XHF=0.4$. The dotted line separates reactions of the \ac{H2}-submechanism from reactions of the \ac{NH3}-submechanism, as the latter are irrelevant for pure \ac{H2} cases. The equivalence ratio is $\phi=0.8$ and temperature is $\Tu=298\rm~K$ for all cases. Positive values indicated an increase of $\mathit{Ze}$ with increasing rate coefficients, hence decreasing reactivity. The opposite holds true for negative sensitivities. The numbering of reactions refers to the order of appearance in the utilized mechanism~\cite{Zhang2021}.}
	\label{fig:Sensitivity}
\end{figure}

The overall reactivity, represented by the Zeldovich number, is closely coupled to chemical kinetics. Therefore, the normalized sensitivity coefficients $\mathcal{S}_{\mathit{Ze},k_i}$ of $\mathit{Ze}$ on the elementary reaction rate coefficient $k_i$ can give insights into the reasons for the pressure dependency of the Zeldovich number\footnote{Note that the reaction rate coefficient $k$ in \cref{eq:Sensitivity} is completely unrelated to the wavenumber $k$ in \cref{eq:Matalon}, despite the same variable name. However, we chose to follow the naming conventions for both quantities.}. The normalized sensitivity coefficients shown in \cref{fig:Sensitivity} are approximated through a central differences brute force method~\cite{Turanyi1990},
\begin{equation}
	\mathcal{S}_{\mathit{Ze},k_i}=\frac{1}{\mathit{Ze}}\frac{\rmd \mathit{Ze}}{\rmd k_i} \approx \frac{1}{\mathit{Ze}} \frac{\mathit{Ze} \left(k_i + \Delta k_i\right) - \mathit{Ze} \left(k_i - \Delta k_i\right)}{2\Delta k_i}\,.
	\label{eq:Sensitivity}
\end{equation}

More specifically, $\mathit{Ze} \left(k_i \pm \Delta k_i\right)$ is the Zeldovich number computed as previously discussed in \cref{sec:TheoreticalBackground}, however, based on a kinetic mechanism where the rate of the $i$-th reaction is perturbed by $\pm \Delta k_i$. In particular, this is achieved by perturbing the pre-exponential factor $A_i$ in the Arrhenius form by 10\%.

\begin{figure}[h!]
	\centering
	\ifdefined\preprint
	\includegraphics[width=0.5\linewidth]{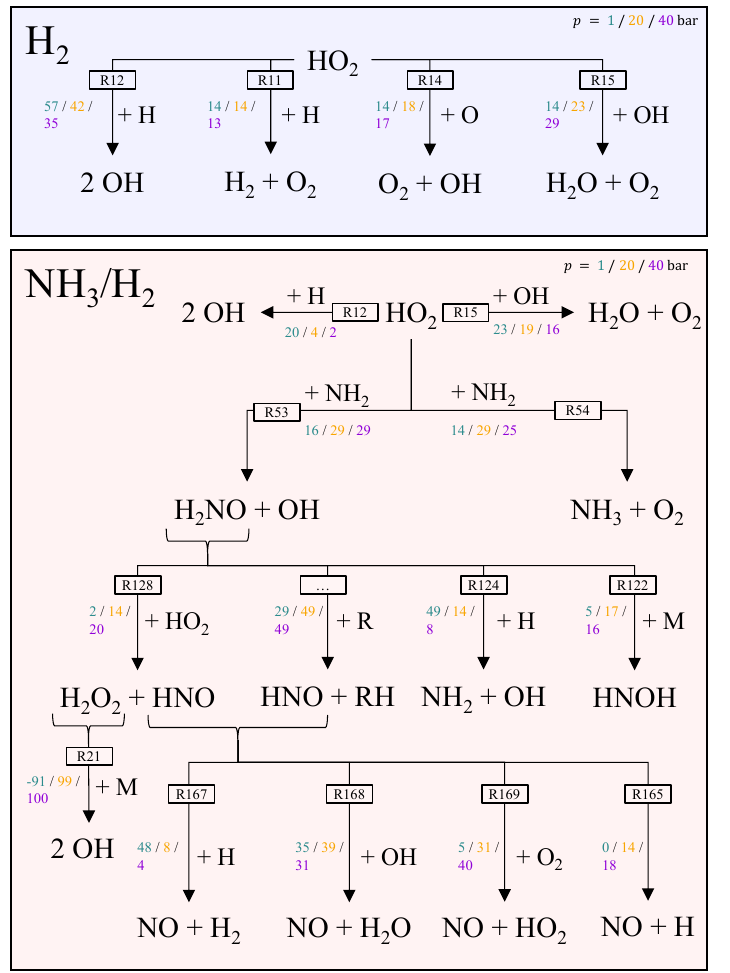}
	\fi
	\ifdefined\final
	\includegraphics[width=\linewidth]{Pathways.pdf}
	\fi
	\caption{Reaction pathway diagram for the reactions of \ce{HO2} for pure \ac{H2} (top) and a \ac{NH3}/\ac{H2} blend with $\XHF=0.4$ (bottom). Numbers represent the relative consumption-based integrated fluxes at 1, 20, and 40 bar in a laminar premixed unstretched flame. The equivalence ratio is $\phi=0.8$ and the temperature is $\Tu=298\rm~K$ for all cases. The path \ce{H2NO + R} lumps multiple hydrogen abstraction reactions via different radicals, and hence no reaction number is assigned. For R21, a negative flux is stated for the low pressure case, indicating a reverse net flux.}
	\label{fig:Pathway}
\end{figure}

As discussed by \citet{Attili2021}, the pressure dependency of the three-body reaction \ce{H + O2 ({+}M) $\leftrightarrow$ HO2 ({+}M)} (R10) plays an important role for the influence of pressure on \acp{IFI} in pure \ac{H2}/air flames. The most important consumption reactions of \ce{HO2} are provided in \cref{fig:Pathway}. Here it becomes evident that R10, together with its subsequent reactions, is net chain terminating in \ac{H2}/air flames, hence leading to decreased reactivity. This is also reflected in the sensitivity coefficients, which increase with increasing pressure. The competing reaction R1: \ce{H + O2 $\leftrightarrow$ O + OH} is chain branching, hence leading to increased overall reactivity. However, as a consequence of pressure dependence of R10, R10 dominates R1 at higher pressure. It should be noted that, although not relevant for the investigated parameter space, at very high pressure or leaner conditions, an increasing importance of \ce{HO2 + H2 $\leftrightarrow$ H2O2 + H} is observed, where \ce{H2O2} reacts to form \ce{2OH} via R21 ~\cite{Howarth2022, Law2006b}. This leads to an increase in reactivity, and hence a decrease of \acp{IFI} as observed by \citet{Howarth2022}. This pressure regime, where \acp{IFI} start to decrease, is outside the parameter space of this study, so that the trend is not visible in the provided data.

For an \ac{NH3}/\ac{H2} blend with $\XHF=0.4$, fluxes towards \ce{HO2} through R10 also increase with $p$. However, the sensitivity of $\mathit{Ze}$ decreases, indicating the vanishing influence of R10 on the overall reactivity. An explanation is given through the pathway analysis shown in \cref{fig:Pathway}. Similar to \ac{H2}/air flames, the reactions R15 and R12 offer consumption pathways for \ce{HO2}, where the latter reaction is only important at low $p$ due to the enhanced consumption of \ce{H} by R10 at high $p$. However, the introduction of \ac{NH3} to the flame opens new pathways through reactions with \ce{NH2}. The reaction \ce{HO2 + NH2 $\leftrightarrow$ NH3 + O2} (R54) is the equivalent to R15, also acting to inhibit, as indicated by the positive $\mathit{Ze}$-sensitivities. On the other hand, the competing reaction \ce{HO2 + NH2 $\leftrightarrow$ H2NO + OH} (R53) acts to accelerate, especially for high $p$. This is mainly related to the reaction \ce{H2NO + HO2 $\leftrightarrow$ H2O2 + HNO} (R205), which becomes more important with increasing \ce{HO2} concentration. As in pure \ac{H2}/air flames, \ce{H2O2} reacts to form \ce{2OH} radicals (R21), more significantly at high $p$, making this channel increasingly chain branching. In total, with the two parallel pathways via R54 and R53, which act to inhibit and accelerate, respectively, the overall effect of the \ce{HO2}-radical becomes chain propagating at high $p$, hence reducing the influence of R10. This leads to a low sensitivity of $\mathit{Ze}$ on $p$ at high $p$.

\subsection{Comparison with theoretical models}
\label{subsec:ComparisonTheoreticalModels}

\begin{figure*}[t]
	\centering
	\includegraphics[width=\textwidth]{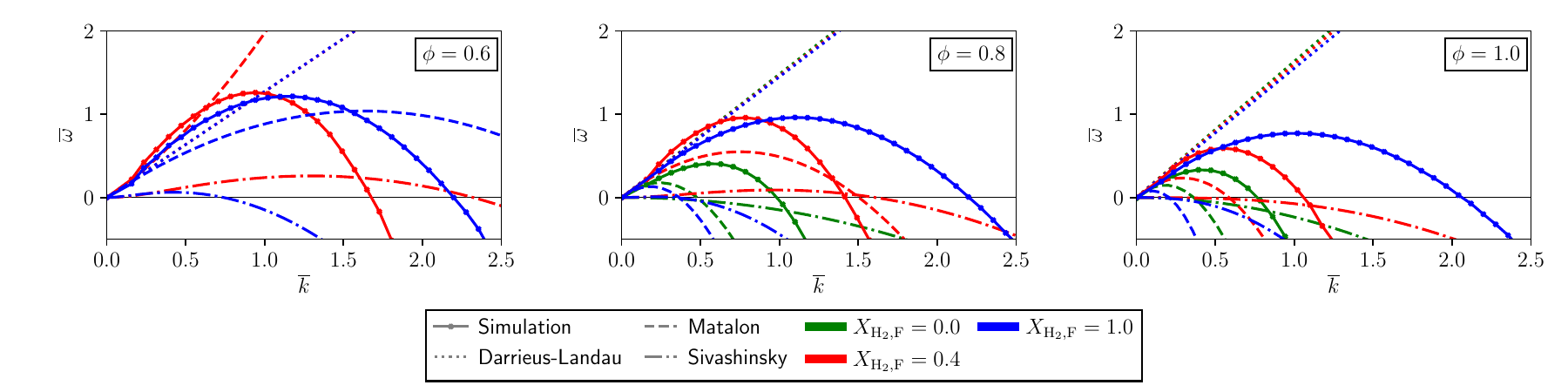}
	\caption{Comparison of numerically calculated dispersion relations with theoretical models: Symbols represent the numerically determined growth rates and solid lines the piecewise linear functions connecting these data. Dotted lines represent the theoretical growth rate of hydrodynamic instabilities, see \cref{eq:omega_0}. Dashed lined represent the theoretical formulation by \citet{Matalon2003}, cf. \cref{eq:Matalon}, and dashed-dotted lines the theoretical formulation by \citet{Sivashinsky1977}, cf. \cref{eq:Sivashinsky}. All simulations are conducted at ambient conditions, $T_{\rm u}=298~\rm K$ and $p=1~\rm bar$.}
	\label{fig:DispRel_Theory}
\end{figure*}

In \cref{fig:DispRel_Theory}, numerically derived dispersion relations are compared with the theoretically derived models by \citet{Matalon2003}, cf. \cref{eq:Matalon}, and \citet{Sivashinsky1977}, cf.~\cref{eq:Sivashinsky}. As already pointed out in the literature~\cite{Attili2021, Berger2022}, the quantitative prediction capability is limited for both formulations. While \citet{Sivashinsky1977} neglected the influence of hydrodynamic instabilities resulting in significantly lower growth rates, \citet{Matalon2003} truncated the derivation after the second-order term, whereas a fourth order term would have a stabilizing effect for thermo-diffusively unstable cases. Nevertheless, the models can be used to evaluate the stabilizing or destabilizing nature of thermo-diffusive processes. In \cref{eq:Matalon}, the onset of thermo-diffusive instabilities is represented by a positive second-order term, and consequently an unconditionally increasing growth rate. In the model by \citet{Sivashinsky1977}, positive growth rates are always a result of thermo-diffusive instabilities. In the lean limit for $\phi=0.6$, both expressions correctly predict the non-monotonic behavior of the growth rate with respect to $\XHF$. For richer mixtures, the non-monotonic behavior is not well captured, hence a prediction of the onset of thermo-diffusive instabilities is not accurate. 

\begin{figure}[h!]
	\centering
	\ifdefined\preprint
	\includegraphics[width=0.5\linewidth]{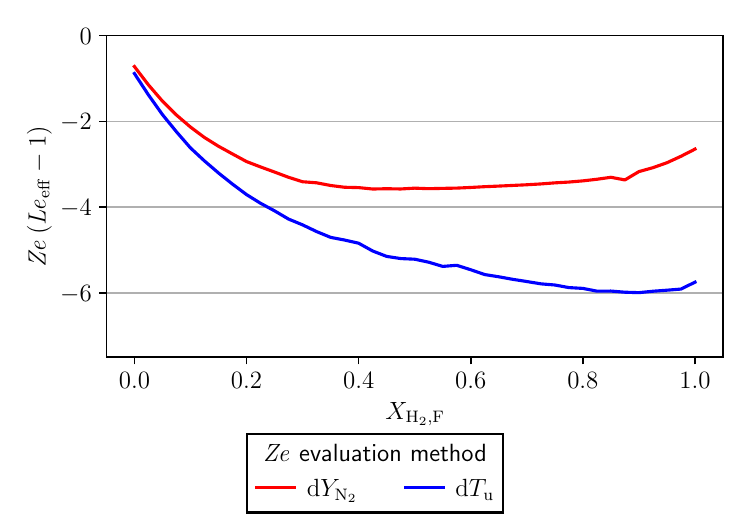}
	\fi
	\ifdefined\final
	\includegraphics[width=\linewidth]{ZeldovichVariation.pdf}
	\fi
	\caption{Influence of different evaluation methods for $\mathit{Ze}$ on $\mathit{Ze}\left(\mathit{Le}-1\right)$, exemplary at $\phi=0.8$: Evaluation of the differential in \cref{eq:ActivationEnergy} through dilution by additional $3~\%$ \ce{N2} (red) or through variation of $T_{\rm u}$ by $\Delta T_{\rm u}=\pm3~\rm K$ (blue).  
	}
	\label{fig:ZeldovichVar}
\end{figure}

At the same time, it should be noted that the predictions are sensitive to $\mathit{Ze}$, and hence to the choice of the numerical evaluation method for the activation energy $E/R$. To understand their influence on the prediction result, the two most common techniques, i.e. based on the dilution of the flame and based on a variation of the unburned temperature, as described in \cref{sec:TheoreticalBackground}, are examined in \cref{fig:ZeldovichVar}. It is observed that the absolute values of $\mathit{Ze}\left(\mathit{Le}-1\right)$ can differ strongly. However, the trends with respect to $\phi$ and $\XHF$ are mostly recovered by any formulation. As a result, the choice of a numerical method can alter the location of the root of $\omega_2$, i.e., the predicted onset of \ac{TDI}, with respect to $\XHF$, while the presented non-monotonic behavior is maintained for any of the tested methods. This underlines the necessity for further analysis of the numerical approaches to determine $\mathit{Ze}$ to result in accurate predictions. Within the scope of this work, the method via dilution was chosen, since the results show the best correlation with the trends observed in the numerically obtained dispersion relations.

\section{Conclusions}
\label{sec:Conclusions}

In this work, \acfp{IFI} in \ac{NH3}/\ac{H2}/air flames have been analyzed in the linear regime with respect to variations of equivalence ratio, \ac{H2} fraction in the fuel blend,  pressure, and unburned temperature. For this purpose, highly resolved two-dimensional direct numerical simulations of statistically planar laminar premixed \ac{NH3}/\ac{H2}/air flames were conducted for various conditions along five parametric variations.

As observed for pure \ac{H2}/air flames, \acp{IFI} are amplified with decreasing equivalence ratio due to an increase of the Zeldovich number and a decrease of the effective Lewis number. Additionally, the effect of thermo-diffusive instabilities features a considerable non-linear, and especially non-monotonic behavior for variations of the fuel blend ratio in \ac{NH3}/\ac{H2}/air flames. This can adequately be represented through the joint effect of Zeldovich number $\mathit{Ze}$ and the reduced effective Lewis number $\mathit{Le}_{\rm eff}$, also appearing in the second order term in the model by \citet{Matalon2003}. At sufficiently lean conditions, \ac{NH3}/\ac{H2} blends with an \ac{H2} fraction of $40\%$ show the strongest thermo-diffusively driven instabilities.

For increasing pressure, \acp{TDI} in \ac{NH3}/\ac{H2} blends first show an increase followed by a moderate decrease. For a blend with 40\% \ac{H2}, this results in an instability peak at $p=10~\rm bar$ for $\phi=0.8$. This can be explained through the non-monotonic behavior of the overall reactivity reflected in the Zeldovich number. Increasing pressure increases the concentration of the hydroperoxyl radical \ce{HO2}, which is consumed via  chain terminating reactions in \ac{H2}/air flames for moderate pressures considered in this study. The addition of \ac{NH3}, however, opens additional reaction pathways, including a net chain branching pathway via \ce{HO2 + NH2 $~\rightarrow~$ H2NO $~\rightarrow~$ H2O2 $~\rightarrow~$ 2OH}, and the parallel chain terminating pathway \ce{NH2 + HO2 $~\rightarrow~$ NH3 + O2}. Globally, these pathways seem to almost balance each other, resulting in an overall almost constant Zeldovich number at higher pressures.

For the comparison of the numerically computed dispersion relations with the theoretically derived models by \citet{Matalon2003} and \citet{Sivashinsky1977}, the focus is set on the qualitative comparison, as the models in their current formulation cannot predict numerical values for thermo-diffusively unstable conditions. However, both formulations can capture the non-monotonic behavior of thermo-diffusively driven instabilities with respect to the \ac{H2} fraction in the fuel blend for lean mixtures. Since this trend is already reflected in the second order term of the model, $\omega_2$, this is not surprising. However, the results are strongly sensitive to the numerical method to determine the Zeldovich number, hence more investigation is needed to determine its appropriate formulation for two-fuel mixtures. 

\section*{Acknowledgements}
\label{sec:Acknowledgements}

TL, TLH, MG, and HP gratefully acknowledge the received funding from the European Research Council (ERC) under the European Union’s Horizon 2020 research and innovation program (Grant agreement No. 101054894). The work of SG is supported by the Deutsche Forschungsgemeinschaft (DFG, German Research Foundation) under Germany's Excellence Strategy - Cluster of Excellence 2186 “The Fuel Science Center” - ID: 390919832.

The authors gratefully acknowledge the computing time provided to them at the NHR Center NHR4CES at RWTH Aachen University (project numbers p0020340, p0020410). This is funded by the Federal Ministry of Education and Research, and the state governments participating on the basis of the resolutions of the GWK for national high performance computing at universities (www.nhr-verein.de/unsere-partner).

\end{document}


\begin{frontmatter}

\title{Comprehensive linear stability analysis for intrinsic instabilities in premixed ammonia/hydrogen/air flames}

\author[fir]{Terence Lehmann\corref{cor1}}
\author[fir]{Lukas Berger}
\author[fir]{Thomas L. Howarth}
\author[fir]{Michael Gauding}
\author[fir]{Sanket Girhe}
\author[sec]{Bassam B. Dally}
\author[fir]{Heinz Pitsch}

\address[fir]{Institute for Combustion Technology, RWTH Aachen University, Templergraben 64, 52056 Aachen, Germany}
\address[sec]{Clean Energy Research Platform, King Abdullah University of Science and Technology, Thuwal, 23955-6900, Saudi-Arabia}
\cortext[cor1]{Corresponding author: t.lehmann@itv.rwth-aachen.de}

\end{frontmatter}

This supplementary material consists of the following parts:

\begin{compactitem}
	\item Coefficients for the dispersion relation model by \citet{Matalon2003}.
	\item Grid independence study.
	\item Independence study for the variable and value for iso-surface generation. 
	\item Complete list of all investigated conditions, associated flame properties, and dispersion relation results.
	\item Comprehensive collection graphical dispersion representations.
\end{compactitem}

\clearpage
\section{Matalon coefficients}
\label{supp:MatalonCoeffs}

The coefficients $B_1$, $B_2$, and $B_3$ in the model by \citet{Matalon2003} are detailed by \citet{Altantzis2012} as
\begin{eqnarray}
	B_1 & = & \frac{\sigma}{2} \left[\frac{\sigma\left(2\omega_0+\sigma+1\right)}{\left(\sigma-1\right)\left[\sigma+\left(\sigma+1\right)\omega_0\right]}\int_{1}^{\sigma}{\frac{\tilde{\lambda}(x)}{x}\rmd x}+\frac{1}{\sigma + \left(\sigma +1\right)\omega_0}\int_{1}^{\sigma}{\tilde{\lambda}(x)\rmd x}\right]\\
	B_2 & = & \frac{\sigma}{2} \left[\frac{\left(1 +\omega_0\right)\left(\sigma+\omega_0\right)}{\left(\sigma-1\right)\left[\sigma+\left(\sigma+1\right)\omega_0\right]}\int_{1}^{\sigma}{\ln\left(\frac{\sigma-1}{x-1}\right)\frac{\tilde{\lambda}(x)}{x}\rmd x}\right]\\
	B_3 & = & \frac{\sigma}{2} \left[\frac{2\left(\sigma-1\right)}{\sigma+\left(\sigma+1\right)\omega_0}\tilde{\lambda}(\sigma)-\frac{2}{\sigma + \left(\sigma +1\right)\omega_0}\int_{1}^{\sigma}{\tilde{\lambda}(x)\rmd x}\right]\,.
\end{eqnarray}
\noindent
Here, $\sigma$ is the expansion ratio, $\omega_0=\left(\sqrt{\sigma^3+\sigma^2-\sigma}-\sigma\right)/\left(\sigma+1\right)$ the hydrodynamic growth rate, $x=T/T_{\rm u}$ the non-dimensional temperature, and $\tilde{\lambda}=\lambda/\lambda_{\rm u}$ the non-dimensional thermal conduction. Within the scope of this work, $\tilde{\lambda}(x)$ and associated integrals are evaluated numerically from one-dimensional flamelet solutions.
\clearpage
\section{Grid independence study}
\label{supp:GridStudy}

\begin{figure}[h!]
	\centering
	\includegraphics[width=0.5\linewidth]{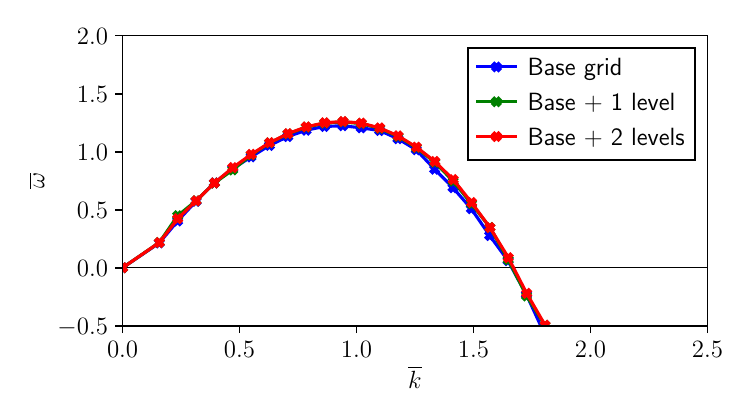}
	\caption{Dispersion relations using different number of refinement levels, i.e. 0 to 2 levels, leading to effective resolutions of $\Delta x = l_{\rm F}/12.8$, $\Delta x = l_{\rm F}/25.6$, and $\Delta x = l_{\rm F}/51.6$.}
	\label{fig:Resolution}
\end{figure}

In order to verify the independence of the results from the grid chosen for the simulation, a grid independence study was executed. \Cref{fig:Resolution} shows the results for simulations using the base grid with a resolution of $\Delta x = l_{\rm F}/12.8$, one level of static local refinement, i.e., $\Delta x = l_{\rm F}/25.6$, and two levels of refinement, i.e., $\Delta x = l_{\rm F}/51.6$. The results clearly show the convergence for the grid with only one level of refinement. In the study, two levels of refinement are applied to ensure the generality for other cases.

\begin{figure}[h!]
	\centering
	\includegraphics[width=0.5\linewidth]{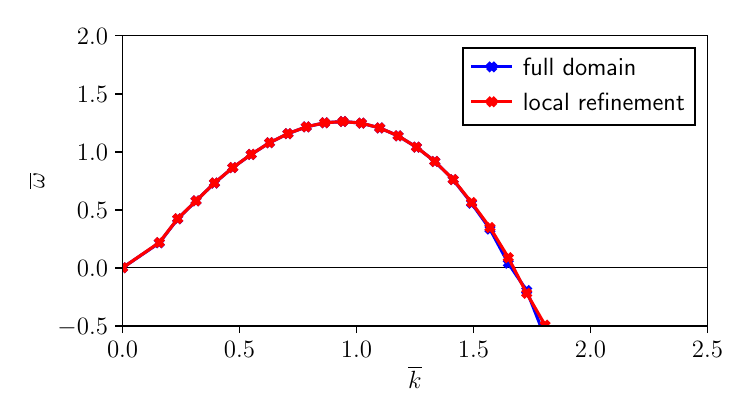}
	\caption{Dispersion relations using the resolution (base + 2 levels) on the whole domain (blue) or only locally (red).}
	\label{fig:Refinement}
\end{figure}

Additionally, the effect of a grid with local refinement compared to a grid with full resolution on the whole domain was analyzed. The dispersion relations presented in \cref{fig:Refinement} show the independence of the results on the local refinement.

\clearpage
\section{Independence study for the variable and value for iso-surface generation}
\label{supp:Isolines}

\begin{figure}[h!]
	\centering
	
	\begin{subfigure}[b]{0.49\linewidth}
		\centering
		\includegraphics[width=\linewidth]{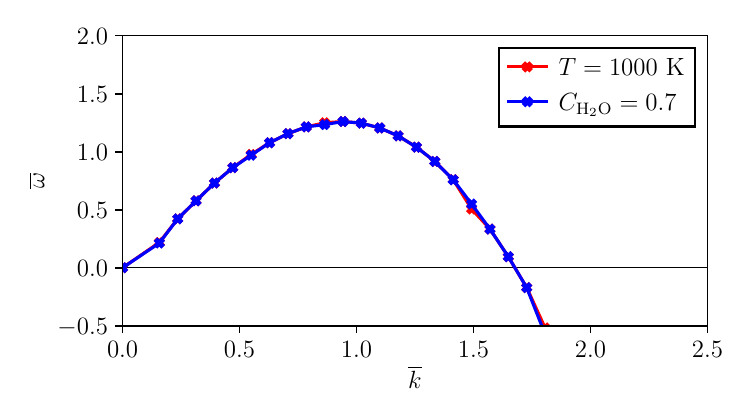}
		\caption{}
		\label{fig:Iso_T_C}
	\end{subfigure}
	\begin{subfigure}[b]{0.49\linewidth}
		\centering
		\includegraphics[width=\linewidth]{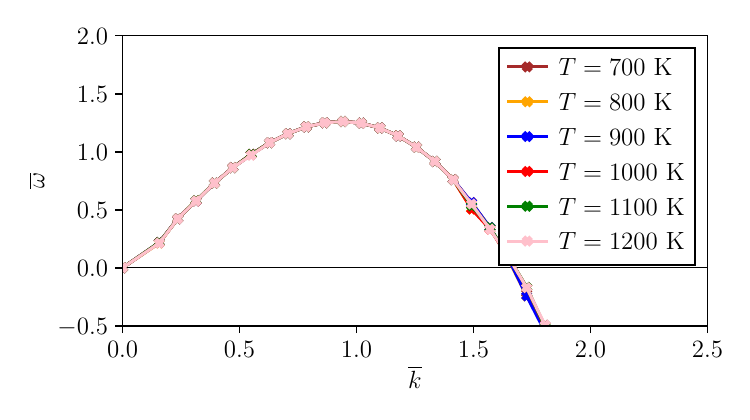}
		\caption{}
		\label{fig:Iso_T_var}
	\end{subfigure}
	
	\begin{subfigure}[b]{\linewidth}
		\centering
		\includegraphics[width=0.5\linewidth]{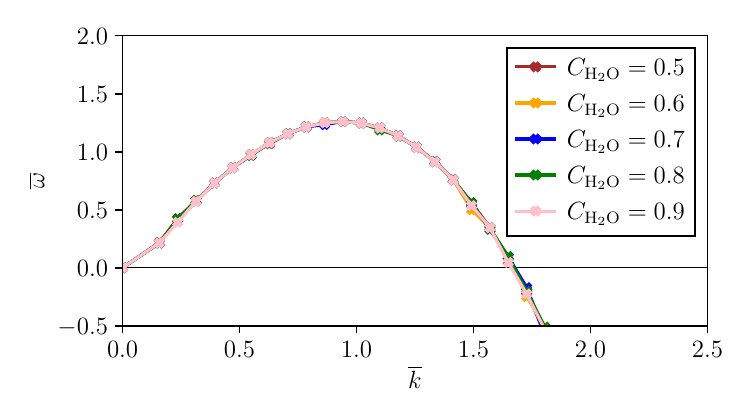}
		\caption{}
		\label{fig:Iso_C_var}
	\end{subfigure}
	\caption{Dispersion relations using different variables and values to determine the iso-surface to trace the flame front. (a) Comparison between $T$- and $C_{\rm H_2O}$-based iso-surface, (b) comparison of different values for definition of $T$-based iso-surface, and (c) comparison of different values for definition of $C_{\rm H_2O}$-based iso-surface. All simulations are conducted at $\phi=0.6$, $\XHF=0.4$, $T_{\rm u}=298~\rm K$, and $p=1~\rm bar$.}
	\label{fig:Iso_Compare}
\end{figure}

In order to verify the independence of the results from the choice of variable to define the iso-surface as well as from the specific numerical value chose for its definition, an independence study has been conducted. Therefore, two variables are selected, namely the local temperature $T$ and the normalized progress based on water (\ce{H2O}). The latter is defined as
\begin{equation}
	C_{\rm H_2O} = \frac{Y_{\rm H_2O}}{Y_{\rm H_2O, b}}\,,
\end{equation}
\noindent
where $Y_{\rm H_2O, b}$ denotes the equilibrium mass fraction of \ce{H2O} in the burned region. \cref{fig:Iso_Compare} shows a comparison of the two definitions. Furthermore, different numerical values to define the iso-surface are compared. No significant deviation is observed. 

It should be noted that there are many other ways to define the iso-surface, e.g. through a temperature-based progress variable or the density field. However, the results above clearly show the marginal influence of this choice on the results, as long as the variable is well defined and the specific iso-value is located inside the flame zone.
\clearpage
\section{Investigated Conditions}
\label{supp:Cases}

\cref{tab:cases} details the conditions examined within the scope of this work. It further includes the respective flame thicknesses $l_{\rm F}$, burning velocities $s_{\rm L}$, and flame times $\tau_{\rm F}$, expansion ratio $\sigma$, effective Lewis number $\mathit{Le}_{\rm eff}$, and Zeldovich number $\mathit{Ze}$. These values are calculated based on simulations of one-dimensional laminar unstretched premixed flames conducted with FlameMaster~\cite{Pitsch1998}. Additionally, the maximum growth rate $\overline{\omega}_{\rm max}$, the wave number $\overline{k}_{\rm max} = \overline{k}(\overline{\omega}=\overline{\omega}_{\rm max})$ at this growth rate, and the cut-off wave number $\overline{k}_{\rm c} = \overline{k}(\overline{\omega}=0) $ derived from the dispersion relations are provided.

\footnotesize{
 \begin{longtable}[c]{ c c c c c c c c c c c c }
	\caption{Investigated conditions and associated flame properties.\label{tab:cases}}\\
	\hline
	$T_{\rm u}$ & $p$         & $\phi$ & $X_{\rm \ce{H2},F}$ & $s_{\rm L}$           & $l_{\rm F}$           & $\sigma$ & $\mathit{Le}_{\rm eff}$ & $\mathit{Ze}$ & $\overline{k}_{\rm max}$ & $\overline{\omega}_{\rm max}$ & $\overline{k}_{\rm c}$\\
	$[\rm K]$   & $[\rm bar]$ & $[-]$  & $[-]$               & $[\rm \mathrm{cm}/s]$ & $[\rm \mathrm{\mu m}]$ & $[-]$       & $[-]$       & $[-]$  & $[-]$ & $[-]$ & $[-]$ \\
	\hline
	\hline
	\endfirsthead
	
	\multicolumn{12}{c}{\textit{Continuation of Table \ref{tab:cases}}}\\
	\hline
	\hline
	$T_{\rm u}$ & $p$         & $\phi$ & $X_{\rm \ce{H2},F}$ & $s_{\rm L}$           & $l_{\rm F}$           & $\sigma$ & $\mathit{Le}_{\rm eff}$ & $\mathit{Ze}$ & $\overline{k}_{\rm max}$ & $\overline{\omega}_{\rm max}$ & $\overline{k}_{\rm c}$\\
$[\rm K]$   & $[\rm bar]$ & $[-]$  & $[-]$               & $[\rm \mathrm{cm}/s]$ & $[\rm \mathrm{\mu m}]$ & $[-]$       & $[-]$       & $[-]$ & $[-]$ & $[-]$ & $[-]$ \\
	\hline
	\endhead
	
	\hline
	\multicolumn{12}{ c }{\textit{Continued on next page}}\\
	\endfoot
	
	\hline
	\endlastfoot
	
	298 & 1 & 0.4 & 0.4 & 0.84 & 9809.51 & 4.33 & 0.51 & 29.22 & 0.40 & 1.16 & 2.07 \\
	298 & 1 & 0.4 & 0.6 & 3.61 & 2479.50 & 4.33 & 0.43 & 20.95 & 0.60 & 1.14 & 2.02 \\
	298 & 1 & 0.4 & 0.8 & 9.22 & 1140.83 & 4.32 & 0.38 & 14.47 & 0.80 & 1.10 & 1.99 \\
	298 & 1 & 0.4 & 1 & 20.67 & 666.39 & 4.35 & 0.34 & 11.56 & 1.00 & 1.17 & 2.12 \\
	298 & 1 & 0.5 & 0.2 & 1.14 & 7986.53 & 4.94 & 0.66 & 24.23 & 0.20 & 0.95 & 1.73 \\
	298 & 1 & 0.5 & 0.4 & 4.50 & 2188.48 & 4.93 & 0.53 & 17.24 & 0.40 & 0.99 & 1.83 \\
	298 & 1 & 0.5 & 0.6 & 11.70 & 976.03 & 4.90 & 0.46 & 12.94 & 0.60 & 1.01 & 1.80 \\
	298 & 1 & 0.5 & 0.8 & 22.68 & 625.95 & 4.88 & 0.41 & 10.27 & 0.80 & 1.03 & 1.88 \\
	298 & 1 & 0.5 & 1 & 47.35 & 446.47 & 4.94 & 0.38 & 8.62 & 1.00 & 1.15 & 2.14 \\
	298 & 1 & 0.6 & 0.2 & 3.44 & 2997.82 & 5.49 & 0.69 & 15.91 & 0.20 & 0.86 & 1.59 \\
	298 & 1 & 0.6 & 0.4 & 9.69 & 1192.84 & 5.46 & 0.56 & 12.68 & 0.40 & 0.94 & 1.68 \\
	298 & 1 & 0.6 & 0.6 & 21.39 & 653.62 & 5.42 & 0.49 & 10.26 & 0.60 & 0.93 & 1.69 \\
	298 & 1 & 0.6 & 0.8 & 38.78 & 472.73 & 5.39 & 0.45 & 8.76 & 0.80 & 0.94 & 1.82 \\
	298 & 1 & 0.6 & 1 & 79.15 & 378.49 & 5.47 & 0.42 & 7.05 & 1.00 & 1.15 & 2.20 \\
	298 & 1 & 0.8 & 0 & 3.91 & 2912.12 & 6.50 & 0.95 & 13.74 & 0.00 & 0.47 & 0.89 \\
	298 & 1 & 0.8 & 0.2 & 8.53 & 1485.64 & 6.45 & 0.75 & 11.73 & 0.20 & 0.70 & 1.28 \\
	298 & 1 & 0.8 & 0.4 & 20.05 & 737.57 & 6.39 & 0.65 & 10.01 & 0.40 & 0.78 & 1.42 \\
	298 & 1 & 0.8 & 0.6 & 41.64 & 460.18 & 6.31 & 0.59 & 8.54 & 0.60 & 0.80 & 1.48 \\
	298 & 1 & 0.8 & 0.8 & 74.81 & 373.37 & 6.27 & 0.55 & 7.51 & 0.80 & 0.90 & 1.72 \\
	298 & 1 & 0.8 & 1 & 144.25 & 341.52 & 6.33 & 0.54 & 5.80 & 1.00 & 1.11 & 2.20 \\
	298 & 1 & 0.9 & 0 & 5.16 & 2403.69 & 6.94 & 0.97 & 12.94 & 0.00 & 0.23 & 0.83 \\
	298 & 1 & 0.9 & 0.2 & 10.60 & 1304.80 & 6.88 & 0.80 & 11.36 & 0.20 & 0.56 & 1.23 \\
	298 & 1 & 0.9 & 0.4 & 24.52 & 666.78 & 6.80 & 0.71 & 9.75 & 0.40 & 0.68 & 1.26 \\
	298 & 1 & 0.9 & 0.6 & 51.07 & 424.87 & 6.72 & 0.66 & 8.31 & 0.60 & 0.71 & 1.35 \\
	298 & 1 & 0.9 & 0.8 & 92.42 & 355.61 & 6.65 & 0.62 & 7.38 & 0.80 & 0.83 & 1.63 \\
	298 & 1 & 0.9 & 1 & 173.75 & 337.22 & 6.65 & 0.62 & 5.89 & 1.00 & 1.07 & 2.16 \\
	298 & 1 & 1 & 0 & 6.41 & 2101.52 & 7.31 & 1.02 & 12.99 & 0.00 & 0.34 & 0.77 \\
	298 & 1 & 1 & 0.2 & 12.53 & 1193.54 & 7.23 & 0.89 & 11.66 & 0.20 & 0.47 & 0.95 \\
	298 & 1 & 1 & 0.4 & 28.28 & 624.15 & 7.12 & 0.82 & 9.82 & 0.40 & 0.55 & 1.06 \\
	298 & 1 & 1 & 0.6 & 59.02 & 399.13 & 7.00 & 0.77 & 8.38 & 0.60 & 0.61 & 1.18 \\
	298 & 1 & 1 & 0.8 & 107.82 & 339.51 & 6.89 & 0.74 & 7.35 & 0.80 & 0.76 & 1.49 \\
	298 & 1 & 1 & 1 & 199.93 & 331.92 & 6.86 & 0.73 & 6.33 & 1.00 & 1.01 & 2.06 \\
	298 & 1 & 1.1 & 0 & 7.85 & 1776.20 & 7.29 & 1.10 & 9.87 & 0.00 & 0.26 & 0.16 \\
	298 & 1 & 1.1 & 0.2 & 13.89 & 1110.21 & 7.22 & 1.01 & 9.10 & 0.20 & 0.44 & 0.73 \\
	298 & 1 & 1.1 & 0.4 & 29.78 & 603.29 & 7.11 & 0.95 & 8.05 & 0.40 & 0.41 & 0.82 \\
	298 & 1 & 1.1 & 0.6 & 63.71 & 376.98 & 6.98 & 0.91 & 6.89 & 0.60 & 0.50 & 0.98 \\
	298 & 1 & 1.1 & 0.8 & 119.64 & 319.81 & 6.87 & 0.88 & 5.98 & 0.80 & 0.66 & 1.31 \\
	298 & 1 & 1.1 & 1 & 222.33 & 322.72 & 6.91 & 0.86 & 5.64 & 1.00 & 0.93 & 1.92 \\
	298 & 1 & 1.2 & 0.2 & 13.43 & 1163.23 & 7.14 & 1.09 & 8.72 & 0.20 & 0.34 & 0.67 \\
	298 & 1 & 1.2 & 0.4 & 27.64 & 655.95 & 7.04 & 1.04 & 8.10 & 0.40 & 0.38 & 0.71 \\
	298 & 1 & 1.2 & 0.6 & 63.75 & 377.43 & 6.91 & 1.00 & 6.74 & 0.60 & 0.42 & 0.83 \\
	298 & 1 & 1.2 & 0.8 & 126.87 & 307.11 & 6.79 & 0.97 & 5.54 & 0.80 & 0.57 & 1.14 \\
	298 & 1 & 1.2 & 1 & 240.59 & 311.72 & 6.85 & 0.94 & 4.78 & 1.00 & 0.84 & 1.75 \\
	298 & 1 & 1.4 & 0 & 6.04 & 2215.63 & 6.89 & 1.24 & 12.39 & 0.00 & 0.26 & 0.46 \\
	298 & 1 & 1.4 & 0.2 & 10.98 & 1393.96 & 6.88 & 1.20 & 10.24 & 0.20 & 0.37 & 0.60 \\
	298 & 1 & 1.4 & 0.4 & 21.67 & 836.81 & 6.85 & 1.18 & 8.38 & 0.40 & 0.34 & 0.66 \\
	298 & 1 & 1.4 & 0.6 & 53.51 & 438.60 & 6.75 & 1.15 & 7.43 & 0.60 & 0.33 & 0.66 \\
	298 & 1 & 1.4 & 0.8 & 128.03 & 304.00 & 6.62 & 1.12 & 5.67 & 0.80 & 0.44 & 0.91 \\
	298 & 1 & 1.4 & 1 & 264.90 & 296.57 & 6.67 & 1.07 & 4.05 & 1.00 & 0.69 & 1.47 \\
	298 & 10 & 0.8 & 0.4 & 6.40 & 199.97 & 6.45 & 0.62 & 13.84 & 0.40 & 0.89 & 1.58 \\
	298 & 10 & 0.8 & 0.6 & 14.02 & 102.14 & 6.39 & 0.56 & 12.21 & 0.60 & 0.94 & 1.68 \\
	298 & 10 & 0.8 & 0.8 & 32.36 & 53.90 & 6.33 & 0.52 & 10.33 & 0.80 & 0.95 & 1.72 \\
	298 & 10 & 0.8 & 1 & 88.75 & 29.20 & 6.32 & 0.49 & 8.82 & 1.00 & 0.98 & 1.84 \\
	298 & 20 & 0.8 & 0.2 & 2.33 & 253.32 & 6.50 & 0.73 & 15.82 & 0.20 & 0.67 & 1.19 \\
	298 & 20 & 0.8 & 0.4 & 4.36 & 144.56 & 6.46 & 0.62 & 14.61 & 0.40 & 0.83 & 1.48 \\
	298 & 20 & 0.8 & 0.6 & 9.11 & 75.40 & 6.41 & 0.55 & 13.28 & 0.60 & 0.94 & 1.69 \\
	298 & 20 & 0.8 & 0.8 & 21.59 & 36.70 & 6.35 & 0.51 & 11.66 & 0.80 & 0.99 & 1.78 \\
	298 & 20 & 0.8 & 1 & 63.38 & 16.90 & 6.33 & 0.48 & 10.48 & 1.00 & 0.99 & 1.80 \\
	500 & 1 & 0.5 & 0.5 & 32.06 & 719.85 & 3.22 & 0.51 & 9.43 & 0.50 & 0.86 & 1.58 \\
	500 & 1 & 0.8 & 0 & 12.56 & 1726.26 & 4.17 & 0.95 & 10.74 & 0.00 & 0.37 & 0.71 \\
	500 & 1 & 0.8 & 0.2 & 25.28 & 978.96 & 4.13 & 0.76 & 9.26 & 0.20 & 0.61 & 1.13 \\
	500 & 1 & 0.8 & 0.4 & 56.32 & 536.25 & 4.09 & 0.66 & 7.89 & 0.40 & 0.67 & 1.25 \\
	500 & 1 & 0.8 & 0.6 & 113.24 & 371.10 & 4.04 & 0.61 & 6.78 & 0.60 & 0.71 & 1.37 \\
	500 & 1 & 0.8 & 0.8 & 199.41 & 343.12 & 4.02 & 0.58 & 6.01 & 0.80 & 0.88 & 1.71 \\
	500 & 1 & 0.8 & 1 & 367.26 & 396.83 & 4.04 & 0.57 & 4.78 & 1.00 & 1.20 & 2.42 \\
	500 & 10 & 0.5 & 0.5 & 6.03 & 311.35 & 3.26 & 0.49 & 14.93 & 0.50 & 0.97 & 1.77 \\
	700 & 1 & 0.8 & 0 & 29.21 & 1209.86 & 3.19 & 0.96 & 8.73 & 0.00 & 0.35 & 0.69 \\
	700 & 1 & 0.8 & 0.2 & 58.23 & 699.67 & 3.16 & 0.78 & 7.55 & 0.20 & 0.53 & 0.99 \\
	700 & 1 & 0.8 & 0.4 & 123.97 & 418.68 & 3.12 & 0.68 & 6.29 & 0.40 & 0.60 & 1.13 \\
	700 & 1 & 0.8 & 0.6 & 244.97 & 313.79 & 3.09 & 0.63 & 5.43 & 0.60 & 0.66 & 1.27 \\
	700 & 1 & 0.8 & 0.8 & 424.89 & 319.96 & 3.06 & 0.60 & 4.77 & 0.80 & 0.81 & 1.61 \\
	700 & 1 & 0.8 & 1 & 735.02 & 421.75 & 3.04 & 0.59 & 3.99 & 1.00 & 0.87 & 2.21 \\
	700 & 5 & 0.8 & 0 & 21.53 & 307.44 & 3.21 & 0.96 & 9.29 & 0.00 & 0.34 & 0.65 \\
	700 & 5 & 0.8 & 0.2 & 36.23 & 207.64 & 3.18 & 0.77 & 8.48 & 0.20 & 0.57 & 1.05 \\
	700 & 5 & 0.8 & 0.4 & 68.91 & 128.40 & 3.15 & 0.67 & 7.74 & 0.40 & 0.69 & 1.29 \\
	700 & 5 & 0.8 & 0.6 & 142.97 & 81.40 & 3.11 & 0.61 & 6.86 & 0.60 & 0.75 & 1.42 \\
	700 & 5 & 0.8 & 0.8 & 292.40 & 63.10 & 3.08 & 0.57 & 5.99 & 0.80 & 0.87 & 1.67 \\
	700 & 5 & 0.8 & 1 & 658.72 & 61.50 & 3.07 & 0.57 & 4.70 & 1.00 & 1.08 & 2.17 \\
	700 & 10 & 0.8 & 0 & 17.58 & 183.86 & 3.22 & 0.95 & 9.84 & 0.00 & 0.33 & 0.65 \\
	700 & 10 & 0.8 & 0.2 & 28.22 & 128.34 & 3.19 & 0.76 & 9.08 & 0.20 & 0.56 & 1.04 \\
	700 & 10 & 0.8 & 0.4 & 50.57 & 81.90 & 3.16 & 0.66 & 8.37 & 0.40 & 0.73 & 1.30 \\
	700 & 10 & 0.8 & 0.6 & 102.77 & 49.80 & 3.12 & 0.60 & 7.61 & 0.60 & 0.79 & 1.46 \\
	700 & 10 & 0.8 & 0.8 & 223.06 & 33.70 & 3.08 & 0.56 & 6.71 & 0.80 & 0.87 & 1.65 \\
	700 & 10 & 0.8 & 1 & 567.82 & 27.70 & 3.08 & 0.55 & 5.32 & 1.00 & 1.07 & 2.08 \\
	700 & 20 & 0.6 & 0 & 6.23 & 222.59 & 2.81 & 0.92 & 11.63 & 0.00 & 0.31 & 0.59 \\
	700 & 20 & 0.6 & 0.2 & 10.31 & 145.62 & 2.80 & 0.70 & 11.10 & 0.20 & 0.56 & 1.04 \\
	700 & 20 & 0.6 & 0.4 & 18.95 & 85.90 & 2.78 & 0.58 & 10.52 & 0.40 & 0.74 & 1.36 \\
	700 & 20 & 0.6 & 0.6 & 38.67 & 46.90 & 2.76 & 0.50 & 9.74 & 0.60 & 0.89 & 1.60 \\
	700 & 20 & 0.6 & 0.8 & 86.65 & 25.80 & 2.73 & 0.45 & 8.50 & 0.80 & 0.96 & 1.77 \\
	700 & 20 & 0.6 & 1 & 242.29 & 15.00 & 2.74 & 0.42 & 7.41 & 1.00 & 1.07 & 2.01 \\
	700 & 20 & 0.8 & 0 & 14.18 & 112.18 & 3.22 & 0.95 & 10.27 & 0.00 & 0.37 & 0.57 \\
	700 & 20 & 0.8 & 0.2 & 21.76 & 80.70 & 3.19 & 0.76 & 9.62 & 0.20 & 0.54 & 0.94 \\
	700 & 20 & 0.8 & 0.4 & 36.85 & 53.30 & 3.17 & 0.65 & 8.99 & 0.40 & 0.67 & 1.23 \\
	700 & 20 & 0.8 & 0.6 & 71.24 & 32.30 & 3.13 & 0.59 & 8.37 & 0.60 & 0.79 & 1.47 \\
	700 & 20 & 0.8 & 0.8 & 158.27 & 19.50 & 3.09 & 0.55 & 7.51 & 0.80 & 0.87 & 1.64 \\
	700 & 20 & 0.8 & 1 & 449.22 & 13.00 & 3.07 & 0.53 & 6.30 & 1.00 & 1.01 & 1.94 \\
	700 & 20 & 0.9 & 0 & 17.37 & 98.70 & 3.40 & 0.98 & 10.03 & 0.00 & 0.31 & 0.58 \\
	700 & 20 & 0.9 & 0.2 & 26.21 & 72.60 & 3.37 & 0.81 & 9.53 & 0.20 & 0.46 & 0.88 \\
	700 & 20 & 0.9 & 0.4 & 43.76 & 49.10 & 3.33 & 0.71 & 9.04 & 0.40 & 0.62 & 1.15 \\
	700 & 20 & 0.9 & 0.6 & 84.77 & 30.10 & 3.29 & 0.65 & 8.59 & 0.60 & 0.72 & 1.35 \\
	700 & 20 & 0.9 & 0.8 & 192.19 & 18.40 & 3.24 & 0.62 & 7.72 & 0.80 & 0.80 & 1.52 \\
	700 & 20 & 0.9 & 1 & 545.94 & 12.80 & 3.20 & 0.61 & 6.01 & 1.00 & 0.96 & 1.87 \\
	700 & 20 & 1 & 0 & 20.27 & 91.30 & 3.56 & 1.02 & 10.67 & 0.00 & 0.32 & 0.54 \\
	700 & 20 & 1 & 0.2 & 30.37 & 67.80 & 3.52 & 0.89 & 10.31 & 0.20 & 0.41 & 0.79 \\
	700 & 20 & 1 & 0.4 & 50.33 & 46.20 & 3.48 & 0.81 & 9.91 & 0.40 & 0.52 & 0.99 \\
	700 & 20 & 1 & 0.6 & 97.68 & 28.50 & 3.42 & 0.76 & 9.38 & 0.60 & 0.61 & 1.16 \\
	700 & 20 & 1 & 0.8 & 224.00 & 17.50 & 3.36 & 0.73 & 8.32 & 0.80 & 0.69 & 1.33 \\
	700 & 20 & 1 & 1 & 633.08 & 12.60 & 3.30 & 0.72 & 6.12 & 1.00 & 0.88 & 1.71 \\
	700 & 20 & 1.1 & 0 & 23.61 & 80.70 & 3.56 & 1.10 & 8.94 & 0.00 & 0.25 & 0.44 \\
	700 & 20 & 1.1 & 0.2 & 34.69 & 61.20 & 3.53 & 1.01 & 8.66 & 0.20 & 0.32 & 0.60 \\
	700 & 20 & 1.1 & 0.4 & 56.28 & 42.90 & 3.50 & 0.95 & 8.55 & 0.40 & 0.40 & 0.75 \\
	700 & 20 & 1.1 & 0.6 & 107.62 & 27.00 & 3.45 & 0.91 & 8.38 & 0.60 & 0.47 & 0.90 \\
	700 & 20 & 1.1 & 0.8 & 248.81 & 16.60 & 3.39 & 0.88 & 7.75 & 0.80 & 0.56 & 1.09 \\
	700 & 20 & 1.1 & 1 & 706.39 & 12.20 & 3.32 & 0.85 & 5.57 & 1.00 & 0.77 & 1.53 \\
	298 & 20 & 1 & 0.4 & 7.32 & 101.33 & 7.27 & 0.81 & 13.75 & 0.40 & 0.61 & 1.14 \\
	298 & 20 & 1 & 0.6 & 14.77 & 55.90 & 7.19 & 0.75 & 12.94 & 0.60 & 0.67 & 1.26 \\
	298 & 20 & 1 & 0.8 & 35.93 & 27.70 & 7.07 & 0.71 & 11.23 & 0.80 & 0.71 & 1.34 \\
	298 & 20 & 1 & 1 & 113.05 & 13.30 & 6.97 & 0.69 & 9.35 & 1.00 & 0.78 & 1.50 \\
	500 & 1 & 1 & 0 & 18.05 & 1409.87 & 4.62 & 1.03 & 10.82 & 0.00 & 0.33 & 0.63 \\
	500 & 1 & 1 & 0.2 & 33.60 & 858.38 & 4.57 & 0.90 & 9.59 & 0.20 & 0.44 & 0.85 \\
	500 & 1 & 1 & 0.4 & 72.10 & 487.24 & 4.49 & 0.83 & 7.91 & 0.40 & 0.50 & 0.96 \\
	500 & 1 & 1 & 0.6 & 146.34 & 336.75 & 4.41 & 0.79 & 6.76 & 0.60 & 0.57 & 1.11 \\
	500 & 1 & 1 & 0.8 & 261.72 & 317.94 & 4.34 & 0.76 & 6.09 & 0.80 & 0.73 & 1.46 \\
	500 & 1 & 1 & 1 & 466.19 & 389.80 & 4.28 & 0.75 & 5.49 & 1.00 & 1.05 & 2.21 \\
	700 & 1 & 1 & 0 & 40.31 & 1001.50 & 3.49 & 1.04 & 9.39 & 0.00 & 0.30 & 0.56 \\
	700 & 1 & 1 & 0.2 & 71.41 & 653.09 & 3.44 & 0.91 & 8.15 & 0.20 & 0.36 & 0.76 \\
	700 & 1 & 1 & 0.4 & 146.45 & 398.87 & 3.38 & 0.84 & 6.64 & 0.40 & 0.46 & 0.89 \\
	700 & 1 & 1 & 0.6 & 292.55 & 292.83 & 3.32 & 0.80 & 5.69 & 0.60 & 0.53 & 1.04 \\
	700 & 1 & 1 & 0.8 & 515.70 & 298.48 & 3.27 & 0.77 & 5.54 & 0.80 & 0.68 & 1.36 \\
	700 & 1 & 1 & 1 & 871.00 & 412.43 & 3.19 & 0.76 & 4.58 & 1.00 & 0.96 & 2.06 \\

\end{longtable}

}

\clearpage
\section{Comprehensive collection of dispersion relations}
\label{supp:DispersionRelations}

The following sections shows dispersion relations for all considered cases.


\begin{figure}[h!]
	\centering
	
	\begin{subfigure}[b]{0.49\linewidth}
		\centering
		\includegraphics[width=\linewidth]{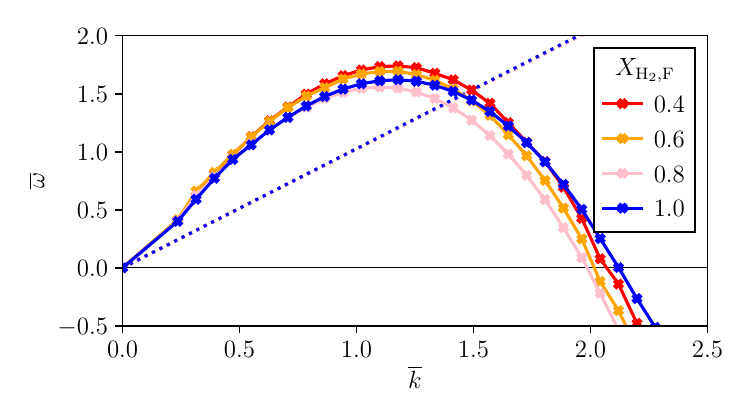}
		\caption{$\phi=0.4$}
		\label{fig:supp_disprel_phi04}
	\end{subfigure}
	\begin{subfigure}[b]{0.49\linewidth}
		\centering
		\includegraphics[width=\linewidth]{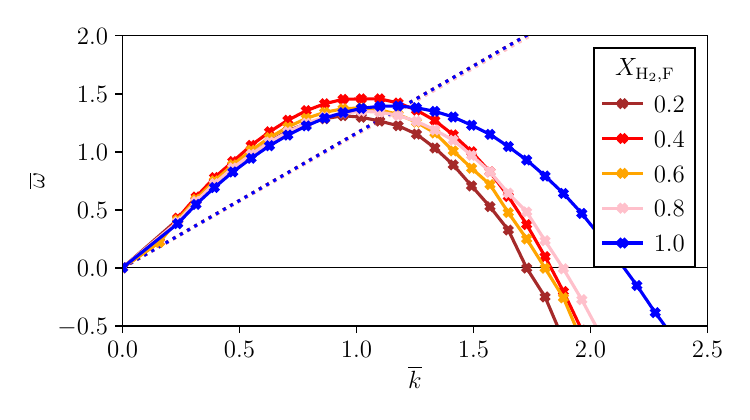}
		\caption{$\phi=0.5$}
		\label{fig:supp_disprel_phi05}
	\end{subfigure}
	
	\begin{subfigure}[b]{0.49\linewidth}
		\centering
		\includegraphics[width=\linewidth]{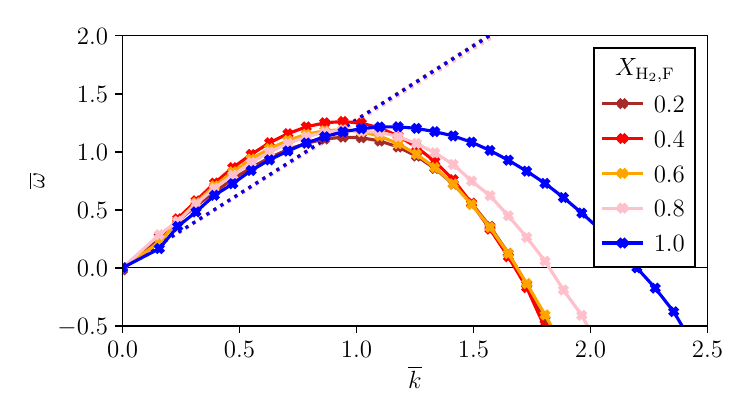}
		\caption{$\phi=0.6$}
		\label{fig:supp_disprel_phi06}
	\end{subfigure}
	\begin{subfigure}[b]{0.49\linewidth}
		\centering
		\includegraphics[width=\linewidth]{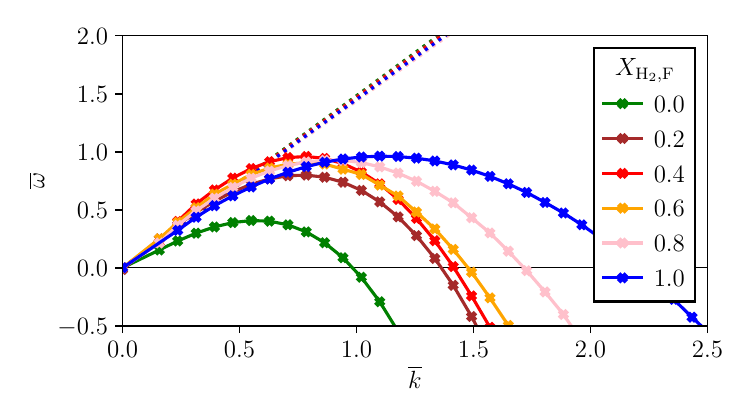}
		\caption{$\phi=0.8$}
		\label{fig:supp_disprel_phi08}
	\end{subfigure}
	
		\begin{subfigure}[b]{0.49\linewidth}
		\centering
		\includegraphics[width=\linewidth]{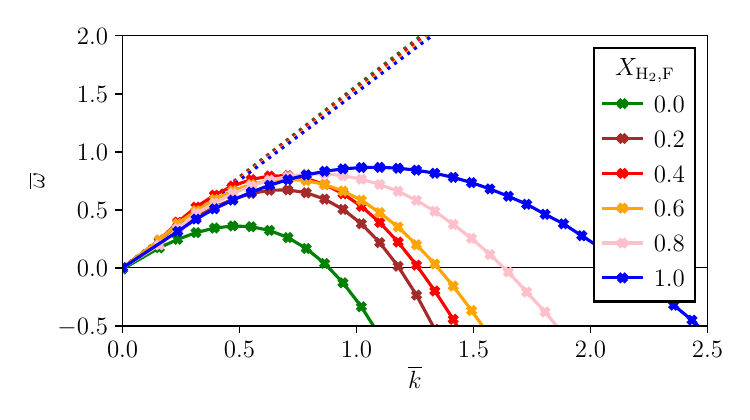}
		\caption{$\phi=0.9$}
		\label{fig:supp_disprel_phi09}
	\end{subfigure}
	\begin{subfigure}[b]{0.49\linewidth}
		\centering
		\includegraphics[width=\linewidth]{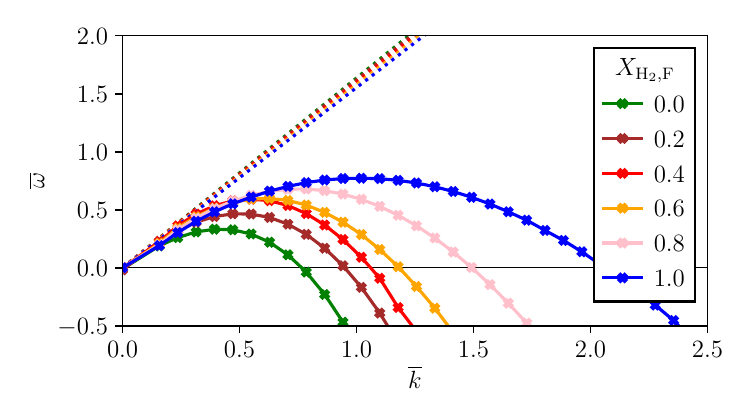}
		\caption{$\phi=1.0$}
		\label{fig:supp_disprel_phi10}
	\end{subfigure}
	
	\caption{Variation 1: Dispersion relations for different molar \ce{H2} contents in the fuel, $X_{\rm H_2,F}$ at different equivalence ratios. All simulations are conducted at $T_{\rm u}=298~\rm K$, and $p=1~\rm bar$.}
\label{fig:supp_disprel_phi}
\end{figure}

\begin{figure}[h!]\ContinuedFloat
	\begin{subfigure}[b]{0.49\linewidth}
		\centering
		\includegraphics[width=\linewidth]{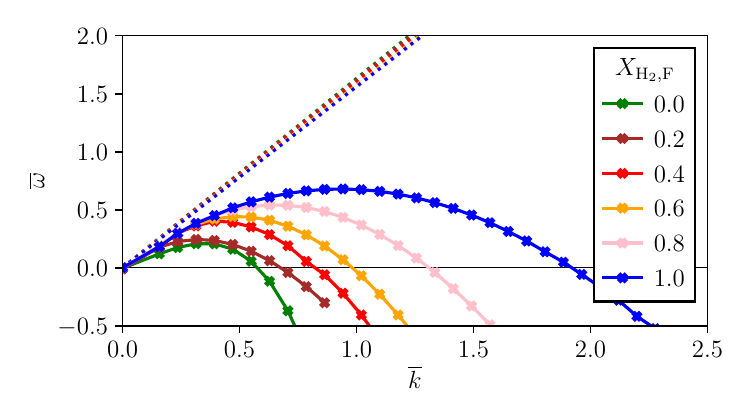}
		\caption{$\phi=1.1$}
		\label{fig:supp_disprel_phi11}
	\end{subfigure}
	\begin{subfigure}[b]{0.49\linewidth}
		\centering
		\includegraphics[width=\linewidth]{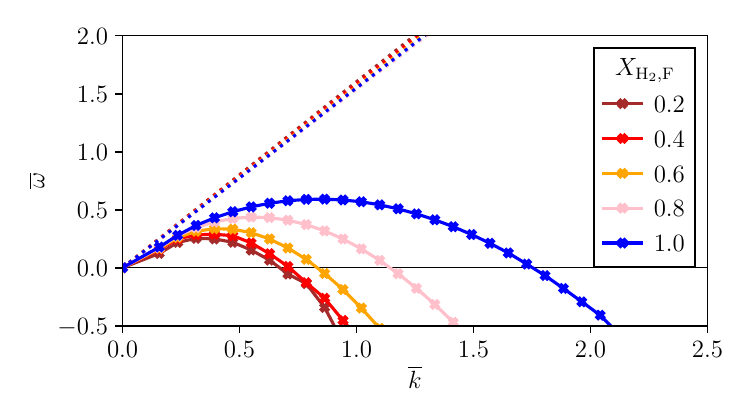}
		\caption{$\phi=1.2$}
		\label{fig:supp_disprel_phi12}
	\end{subfigure}
	
	\begin{subfigure}[b]{\linewidth}
		\centering
		\includegraphics[width=0.5\linewidth]{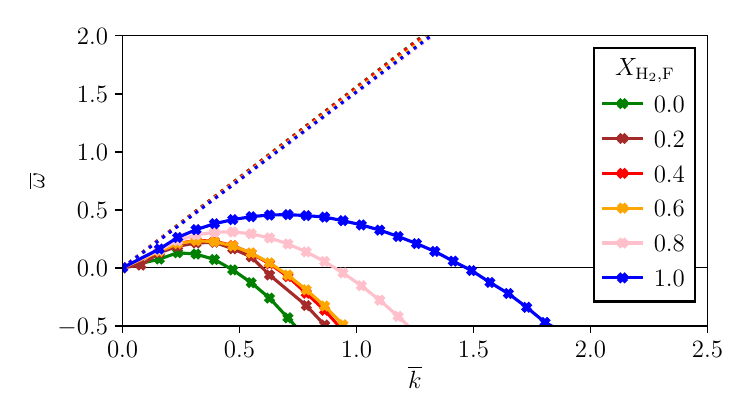}
		\caption{$\phi=1.4$}
		\label{fig:supp_disprel_phi14}
	\end{subfigure}
	\caption[]{(Continued) Variation 1: Dispersion relations for different molar \ce{H2} contents in the fuel, $X_{\rm H_2,F}$ at different equivalence ratios. All simulations are conducted at $T_{\rm u}=298~\rm K$, and $p=1~\rm bar$. }
	\label{fig:}
\end{figure}


\begin{figure}[h!]
	\centering
	
	\begin{subfigure}[b]{0.49\linewidth}
		\centering
		\includegraphics[width=\linewidth]{Figures/SupplementaryMaterial/DispRel_tu0298p01_0phi0_80.pdf}
		\caption{$p=1~\rm bar$}
		\label{fig:supp_disprel_p01}
	\end{subfigure}
	\begin{subfigure}[b]{0.49\linewidth}
		\centering
		\includegraphics[width=\linewidth]{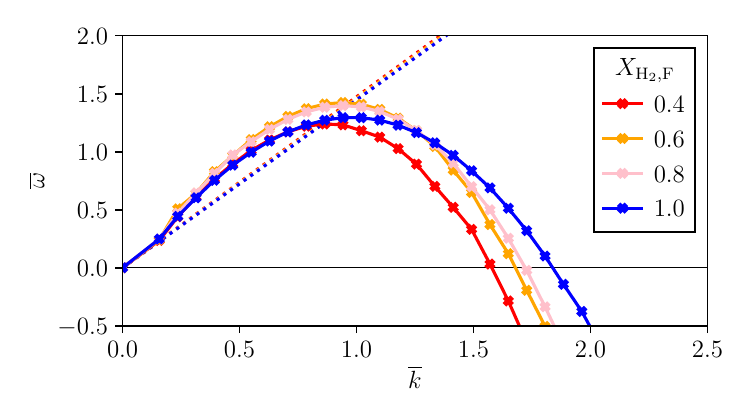}
		\caption{$p=10~\rm bar$}
		\label{fig:supp_disprel_p10}
	\end{subfigure}

	\begin{subfigure}[b]{0.49\linewidth}
		\centering
		\includegraphics[width=\linewidth]{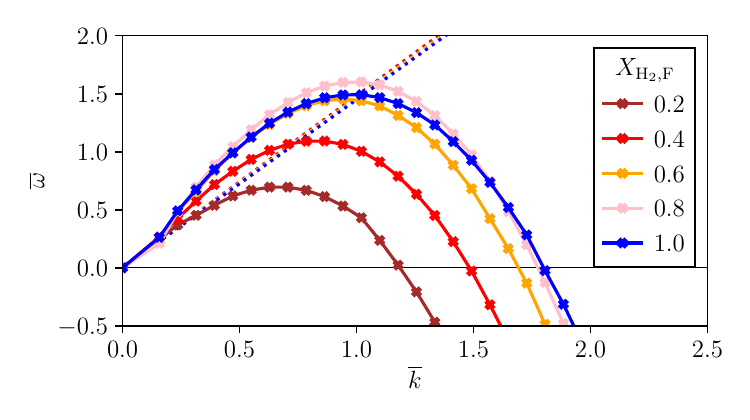}
		\caption{$p=20~\rm bar$}
		\label{fig:supp_disprel_p20}
	\end{subfigure}
	
	\caption{Variation 2: Dispersion relations for different molar \ce{H2} contents in the fuel, $X_{\rm H_2,F}$ at different pressures. All simulations are conducted at $T_{\rm u}=298~\rm K$, and $\phi=0.8$.}
	\label{fig:supp_disprel_p}
\end{figure}


\begin{figure}[h!]
	\centering
	
	\begin{subfigure}[b]{0.49\linewidth}
		\centering
		\includegraphics[width=\linewidth]{Figures/SupplementaryMaterial/DispRel_tu0298p01_0phi0_80.pdf}
		\caption{$T_{\rm u}=298~\rm K$}
		\label{fig:supp_disprel_t298}
	\end{subfigure}
	\begin{subfigure}[b]{0.49\linewidth}
		\centering
		\includegraphics[width=\linewidth]{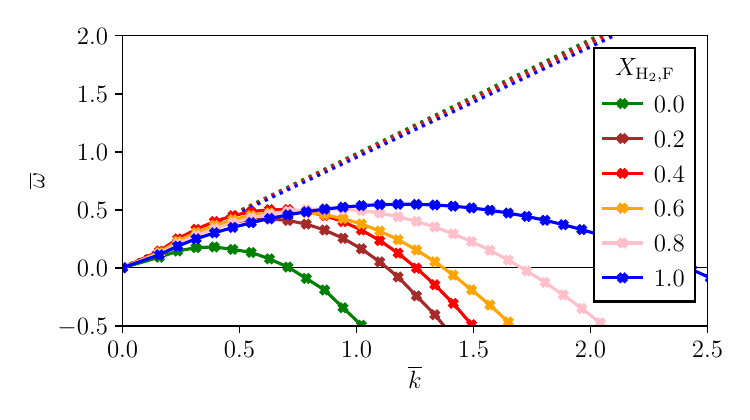}
		\caption{$T_{\rm u}=500~\rm K$}
		\label{fig:supp_disprel_t500}
	\end{subfigure}
	
	\begin{subfigure}[b]{0.49\linewidth}
		\centering
		\includegraphics[width=\linewidth]{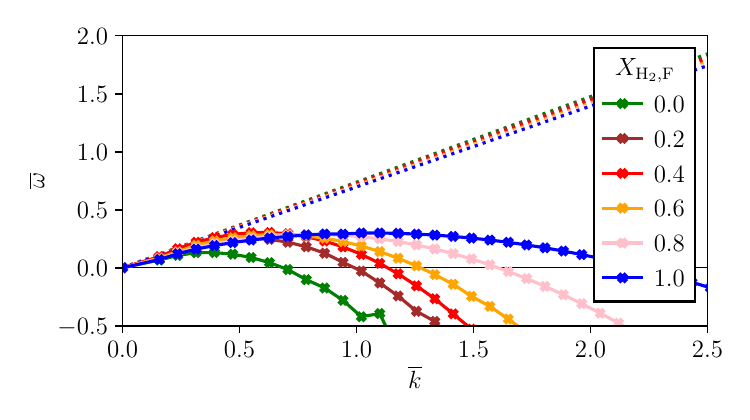}
		\caption{$T_{\rm u}=700~\rm K$}
		\label{fig:supp_disprel_t700}
	\end{subfigure}
	
	\caption{Variation 3: Dispersion relations for different molar \ce{H2} contents in the fuel, $X_{\rm H_2,F}$ at different temperatures. All simulations are conducted at $\phi=0.8$ and $p=1~\rm bar$.}
	\label{fig:supp_disprel_t}
\end{figure}


\begin{figure}[h!]
	\centering
	
	\begin{subfigure}[b]{0.49\linewidth}
		\centering
		\includegraphics[width=\linewidth]{Figures/SupplementaryMaterial/DispRel_tu0700p01_0phi0_80.pdf}
		\caption{$p=1~\rm bar$}
		\label{fig:supp_disprel_t700_p01}
	\end{subfigure}
	\begin{subfigure}[b]{0.49\linewidth}
		\centering
		\includegraphics[width=\linewidth]{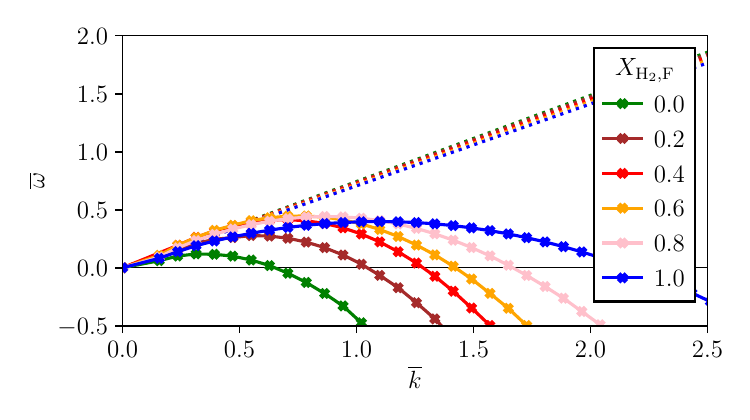}
		\caption{$p=5~\rm bar$}
		\label{fig:supp_disprel_t700_p5}
	\end{subfigure}
	
	\begin{subfigure}[b]{0.49\linewidth}
		\centering
		\includegraphics[width=\linewidth]{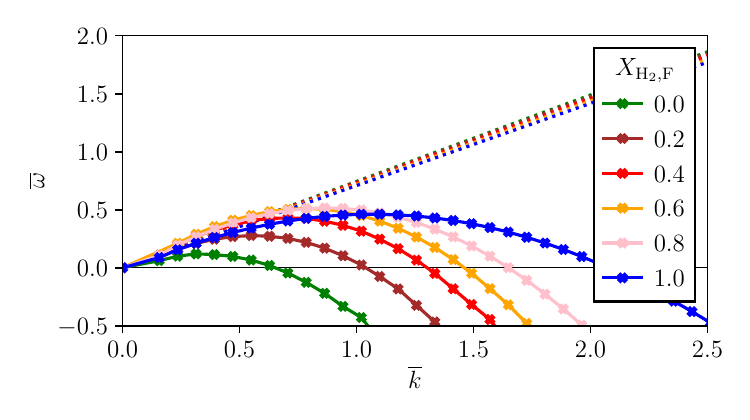}
		\caption{$p=10~\rm bar$}
		\label{fig:supp_disprel_t700_p10}
	\end{subfigure}
		\begin{subfigure}[b]{0.49\linewidth}
		\centering
		\includegraphics[width=\linewidth]{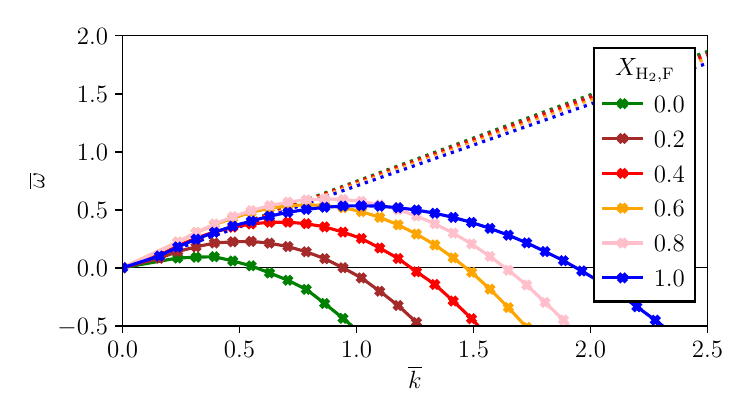}
		\caption{$p=20~\rm bar$}
		\label{fig:supp_disprel_t700_p20}
	\end{subfigure}
	
	\caption{Variation 4: Dispersion relations for different molar \ce{H2} contents in the fuel, $X_{\rm H_2,F}$ at different pressures. All simulations are conducted at $T_{\rm u}=700~\rm K$, and $\phi=0.8$.}
	\label{fig:supp_disprel_p_ht}
\end{figure}


\begin{figure}[h!]
	\centering
	
	\begin{subfigure}[b]{0.49\linewidth}
		\centering
		\includegraphics[width=\linewidth]{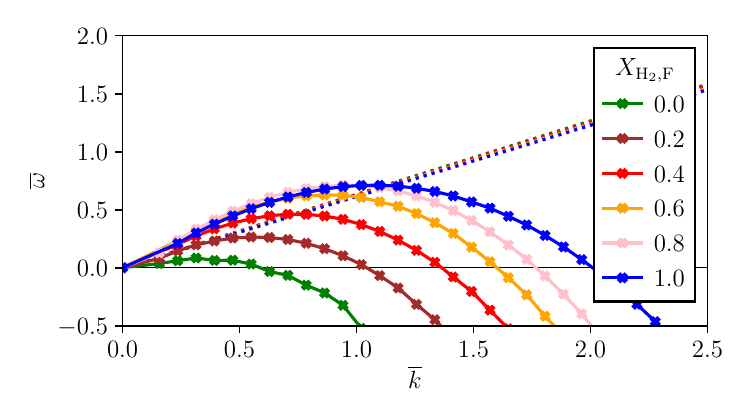}
		\caption{$\phi=0.6$}
		\label{fig:supp_disprel_htp_phi06}
	\end{subfigure}
	\begin{subfigure}[b]{0.49\linewidth}
		\centering
		\includegraphics[width=\linewidth]{Figures/SupplementaryMaterial/DispRel_tu0700p20_0phi0_80.pdf}
		\caption{$\phi=0.8$}
		\label{fig:supp_disprel_htp_phi08}
	\end{subfigure}
	
	\begin{subfigure}[b]{0.49\linewidth}
		\centering
		\includegraphics[width=\linewidth]{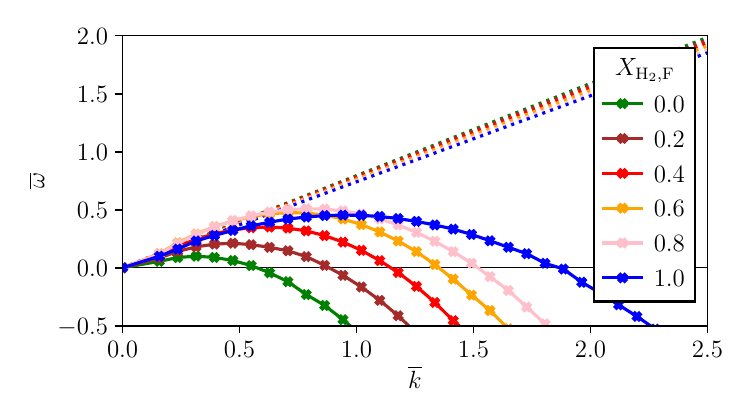}
		\caption{$\phi=0.9$}
		\label{fig:supp_disprel_htp_phi09}
	\end{subfigure}
	\begin{subfigure}[b]{0.49\linewidth}
		\centering
		\includegraphics[width=\linewidth]{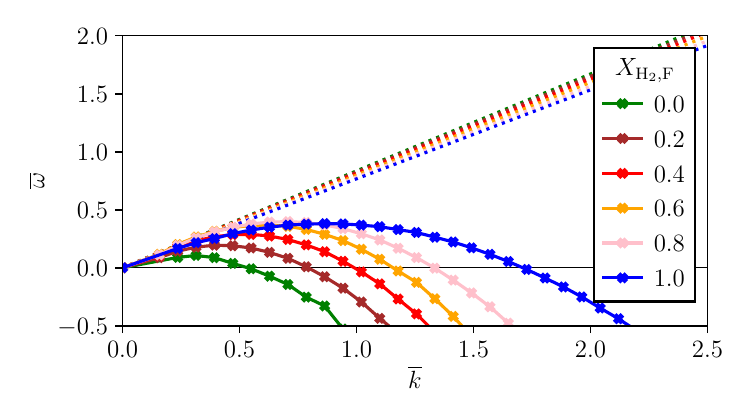}
		\caption{$\phi=1.0$}
		\label{fig:supp_disprel_htp_phi10}
	\end{subfigure}
	
	\begin{subfigure}[b]{0.49\linewidth}
		\centering
		\includegraphics[width=\linewidth]{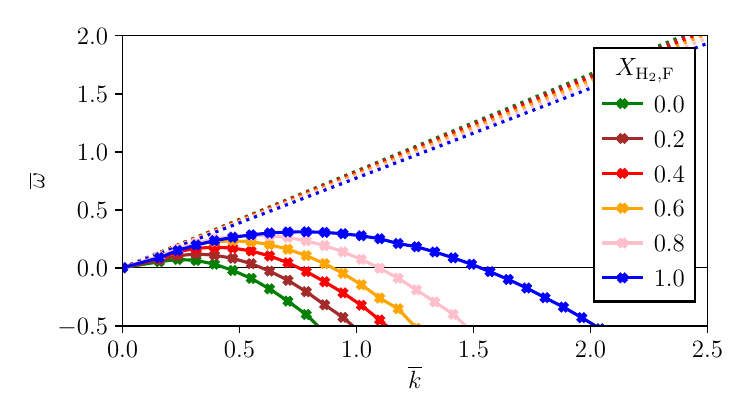}
		\caption{$\phi=1.1$}
		\label{fig:supp_disprel_htp_phi11}
	\end{subfigure}

	\caption{Variation 5: Dispersion relations for different molar \ce{H2} contents in the fuel, $X_{\rm H_2,F}$ at different equivalence ratios. All simulations are conducted at $T_{\rm u}=700~\rm K$, and $p=20~\rm bar$.}
	\label{fig:supp_disprel_dtp_phi}
\end{figure}


\begin{figure}[h!]
	\centering
	
	\begin{subfigure}[b]{0.49\linewidth}
		\centering
		\includegraphics[width=\linewidth]{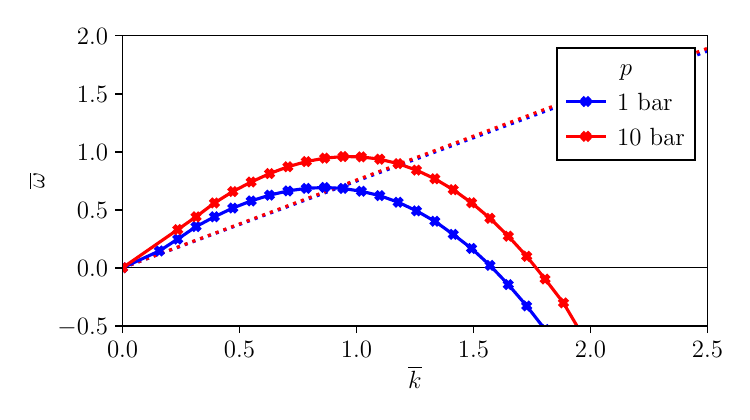}
		\caption{$\phi=0.5$, $T_{\rm u}=500~\rm K$, and $X_{\rm H_2,F}=0.5$}
		\label{fig:supp_disprel_add_1}
	\end{subfigure}
	\begin{subfigure}[b]{0.49\linewidth}
		\centering
		\includegraphics[width=\linewidth]{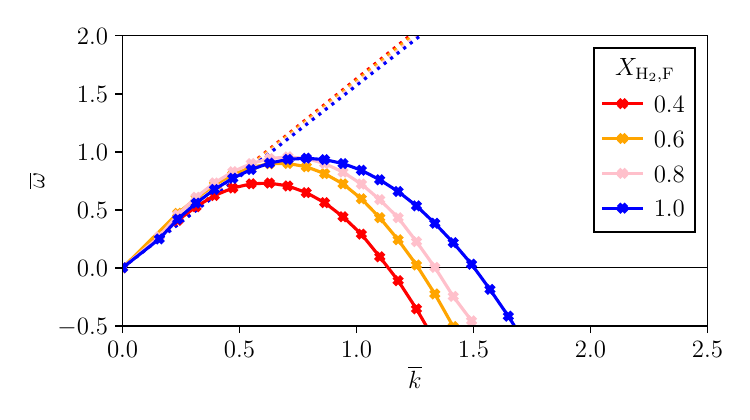}
		\caption{$\phi=1.0$, $T_{\rm u}=298~\rm K$, and $p=20~\rm bar$}
		\label{fig:supp_disprel_add_2}
	\end{subfigure}
	
	\begin{subfigure}[b]{0.49\linewidth}
		\centering
		\includegraphics[width=\linewidth]{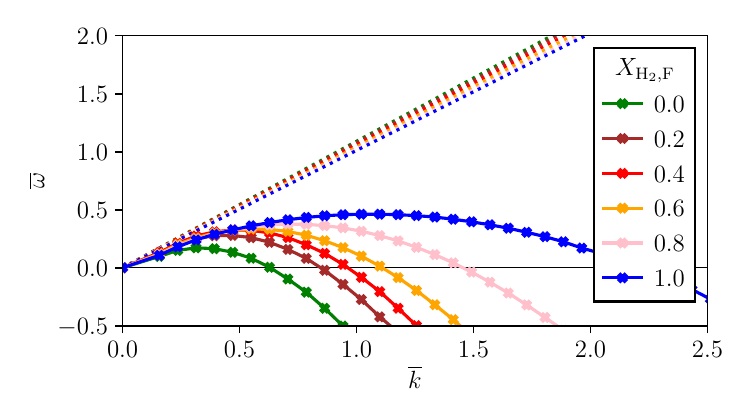}
		\caption{$\phi=1.0$, $T_{\rm u}=500~\rm K$, and $p=1~\rm bar$}
		\label{fig:supp_disprel_add_3}
	\end{subfigure}
	\begin{subfigure}[b]{0.49\linewidth}
		\centering
		\includegraphics[width=\linewidth]{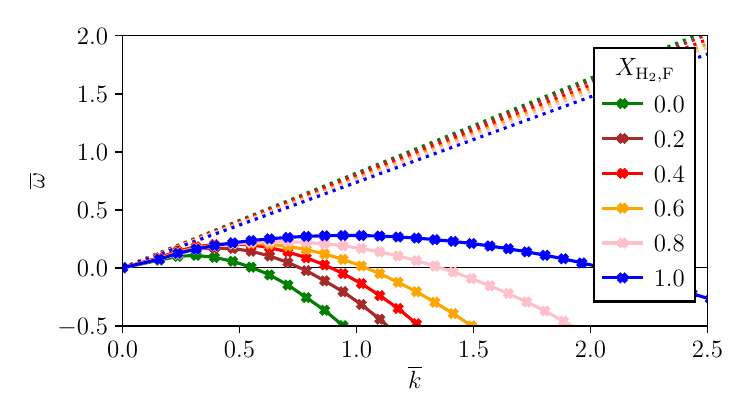}
		\caption{$\phi=1.0$, $T_{\rm u}=700~\rm K$, and $p=1~\rm bar$}
		\label{fig:supp_disprel_add_4}
	\end{subfigure}
	
	\caption{Additional cases: Dispersion relations outside of the presented variations. Conditions are given in the subcaption.}
	\label{fig:supp_disprel_add}
\end{figure}
